\documentclass[twocolumn]{aastex63}

\usepackage{natbib}
\usepackage{color}
\usepackage{mathptmx}
\usepackage{amsmath}
\usepackage{verbatim}
\usepackage{afterpage}  
\usepackage{refcount}
\usepackage{threeparttable}
\usepackage{xcolor}
\usepackage{ulem}

\usepackage{amsmath}	
\usepackage{amssymb}	
\usepackage{url}
\usepackage{multirow}
\usepackage{physics}
\usepackage{amssymb}
\usepackage{pifont}
\newcommand{\cmark}{\ding{51}}%
\newcommand{\xmark}{\ding{55}}%

\usepackage{kotex}
\newcommand{\msun}{\,\rm M_\odot}
\definecolor{ejcol}{rgb}{1,0,0.8}
\definecolor{ejcol2}{rgb}{0,0.8,0.8}

\received{\today}
\revised{\today}
\accepted{February XX, 2021}
\submitjournal{ApJ}

\shorttitle{How Metals Are Transported In And Out Of A Galactic Disk In Simulations}
\shortauthors{Eun-jin Shin, Ji-hoon Kim \& Boon Kiat Oh}
\graphicspath{{./}{figures/}}

\begin{document}

\title{How Metals Are Transported In And Out Of A Galactic Disk: Dependence On The Hydrodynamic Schemes \\  In Numerical Simulations}

\correspondingauthor{Ji-hoon Kim}
\email{me@jihoonkim.org}

\author[0000-0002-4639-5285]{Eun-jin Shin}
\affiliation{Center for Theoretical Physics, Department of Physics and Astronomy, Seoul National University, Seoul 08826, Korea}

\author[0000-0003-4464-1160]{Ji-hoon Kim}
\affiliation{Center for Theoretical Physics, Department of Physics and Astronomy, Seoul National University, Seoul 08826, Korea}
\affil{Seoul National University Astronomy Research Center, Seoul 08826, Korea}

\author[0000-0003-4597-6739]{Boon Kiat Oh}
\affiliation{Center for Theoretical Physics, Department of Physics and Astronomy, Seoul National University, Seoul 08826, Korea}

\begin{abstract}

Metallicity is a fundamental probe for understanding the baryon physics in a galaxy. 
Since metals are intricately associated with radiative cooling, star formation and feedback, reproducing the observed metal distribution through numerical experiments will provide a prominent way to examine our understandings of galactic baryon physics. 
In this study, we analyze the dependence of the galactic metal distribution on the numerical schemes and quantify the differences in the metal mixing among modern galaxy simulation codes (the mesh-based code {\sc Enzo} and the particle-based codes {\sc Gadget-2} and {\sc Gizmo-PSPH}). 
In particular, we examine different stellar feedback strengths and an explicit metal diffusion scheme in particle-based codes, as a way to alleviate the well-known discrepancy in metal transport between mesh-based and particle-based simulations. 
We demonstrate that a sufficient number of gas particles are needed in the gas halo to properly investigate the metal distribution therein.
Including an explicit metal diffusion scheme does not significantly affect the metal distribution in the galactic disk but does change the amount of low-metallicity gas in the hot-diffuse halo. 
We also find that the spatial distribution of metals depends strongly on how the stellar feedback is modeled.
We demonstrate that the previously reported discrepancy in metals between mesh-based and particle-based simulations can be mitigated with our proposed prescription, enabling these simulations to be reliably utilized in the study of metals in galactic halos and the circumgalactic medium. 

\end{abstract}



\section{Introduction}\label{sec:intro}

\vspace{1mm}

Metals in galaxies are fundamental probes to understand the baryon physics in the galactic ecosystem.  
Since metals are mainly produced in stars and released to the interstellar medium (ISM) via supernova (SN) explosion, the metal distribution provides important information on the stellar lifecycle and stellar feedback.
Furthermore, metals are not only the passive tracers of the stellar feedback but also one of the main coolants --- e.g., C and O lines in the warm and neutral component of the ISM \citep{Dalgarno1972} ---  thus playing a crucial role in star formation.
Indeed, the chemical enrichment of a galaxy is the consequence of an intricate interplay of baryonic processes in galaxies, including star formation, gas inflow and outflow, and turbulence. 
Therefore, the metal distribution offers crucial constraints on the processes of galactic evolution.

Observationally, the metal abundance of a galaxy is determined by the flux ratio of line emissions induced by young OB starlight \citep{Tremonti2004, Nagao2006, Liu2008} or by the diffuse ionized gas \citep[DIG;][]{Haffner1999, Sanders2017, Kumari2019}. 
Observations have also revealed a tight global correlation between the stellar mass and gas-phase metallicity ($Z$; defined as a mass fraction of metals over total gas) such that more massive galaxies are more metal-enriched throughout a wide range of galaxy masses and redshifts \citep[MZR;][]{Lequeux1979, Tremonti2004, Mannucci2010, Sanchez2019}.
Moreover, it has been found that the scatter of this relation decreases when star formation rates (SFRs) are also considered \citep{Mannucci2010}, giving rise to the so-called fundamental metallicity relation (FMR) --- a plane $Z = Z\,({\rm SFR}, M_\star)$ in the three-dimensional space of metallicity ($Z$), stellar mass ($M_\star$) --- covering a wide mass range down to dwarf galaxies \citep{Lopez2010, Yates2012,Cresci2019}.
Meanwhile, spatially resolved ISM observations have revealed a negative radial metallicity gradient; in other words, metals are more abundant in the central region than the outer region \citep{Zaritsky1994, Swinbank2012, Jones2013}. 
This gradient can be explained by the inside-out disk growth that produces a negative stellar population gradient \citep{Matteucci1989, Boissier1999,Prantzos2000}.
However, a positive or little gradient has also been reported \citep{Cresci2010, Troncoso2014, Carton2018}.
In particular, interacting galaxies may exhibit lower metallicity in the central region due to the tidally-induced inflow of primordial gas \citep{Cresci2010,Perez2011,Troncoso2014}.

In order to test modern theories of galactic baryon physics and metal transport, numerical experiments have been widely used.  
Simulations have demonstrated that the observed correlation between stellar mass and metallicity can arise naturally in a hierarchical structure formation scenario when the stellar mass growth is regulated by its own negative feedback or mergers \citep[e.g.,][]{Rossi2007, Dave2011}.
The SN-driven outflows are shown to play a key role, especially in low-mass galaxies, as metal-enriched material can escape the galactic potential wells more efficiently, keeping their metallicity low \citep[e.g.,][]{Larson1974, Tremonti2004, Brooks2007, Rossi2017}.
Simulations also found that the galactic environment can alter metal distribution.
For example, the inflow of the pristine gas dilutes the metal distribution in disks \citep[e.g.,][]{Finlator2008, Dave2011}; ram pressure can be exerted by the surrounding gas on to the metal-enriched outflow resulting in the ejecta contained within the galactic halo for an extended amount of time  \cite[e.g.,][]{Ferrara2005}. 

Hydrodynamic simulations have also been utilized to investigate more detailed properties related to metals, such as the metal diffusion coupled with the ISM turbulence and the metallicity distribution function (MDF), which are essential to understand the contamination of pristine gas \citep{Pan2013}.
As for reproducing the metal mixing in simulations, however, both particle-based codes and mesh-based codes have their own issues.
In Lagrangian particle-based schemes such as the smoothed particle hydrodynamics (SPH) approach, metals do not mix between particles unless an explicit diffusion physics is included as a {\it subgrid} model.
On the other hand, Eulerian mesh-based codes inherently allow metals to diffuse. 
However, the artificial diffusion required for stable hydrodynamical solutions may over-mix fluids in simulations with insufficient resolution \citep[e.g.,][]{Pan2013,Springel2016}, especially in systems moving relative to the grid \citep[e.g.,][]{Pontzen2020}.
Naturally, several authors have reported different methods to mitigate the weaknesses in these numerical approaches.  
For particle-based codes, the turbulent diffusion scheme --- calculated with velocity dispersion \citep[e.g.,][]{Greif2009, Revaz2016} or velocity shear \citep[e.g.,][]{Shen2010, Brook2012, Su2017, Escala2018} following \citet{Smagorinsky1963} --- are widely used to estimate the effects of subgrid diffusion and to investigate the MDF (see also \cite{Hu2020} that took a different approach).
Some authors smooth the metallicity over the SPH kernel when they compute, e.g., the cooling rates for gas particles, not actually performing but mimicking the diffusion of metals \citep{Wiersma2009}.  
For mesh-based codes, the over-mixing in unresolved eddies has been addressed with a turbulent diffusion model based on the probability distribution function (PDF) method \citep[e.g.,][]{Pan2013, Sarmento2017}, a different subgrid model based on a partial differential equation for energy density \citep[e.g.,][]{Schmidt2014, Schmidt2015}, or a ``velocity-zeroed'' initial conditions \citep{Pontzen2020}. 

Since metal distribution is highly sensitive to the hydrodynamic scheme and the diffusion model, careful attention on these numerical methods is imperative for studying chemo-dynamical processes in a galaxy using numerical simulations.
The numerical galaxy formation community has collectively responded to this need over the years, and one such effort was the code comparison project {\it AGORA} \citep{AGORA2014, AGORA2016}.
The {\it AGORA} Collaboration tested the reproducibility of numerical experiments using common initial conditions for a dark matter-only cosmological simulation \citep{AGORA2014} and an idealized, isolated galaxy \citep{AGORA2016}, providing insights into both the similarities and differences between contemporary simulation codes.
While reporting solid convergence in many galactic properties, \cite{AGORA2016} also pointed out a discrepancy in the metal distribution between mesh-based and particle-based codes in a test with an idealized disk galaxy.
For example, the metal content of the hot-diffuse halo gas is captured in mesh-based codes, whereas, by design, gas particles are scarce in the halo in particle-based simulations. 
With neither the halo gas nor an explicit metal mixing scheme included \citep[a design choice in][]{AGORA2016}, metal-enriched gas particles tend to stay only near the dense star-forming regions in particle-based simulations \citep[see Figures 32 and 33 in][]{AGORA2016}.

\begin{deluxetable*}{p{1.75cm}p{3.2cm}p{2.9cm}p{2.9cm}p{2.9cm}p{2.9cm}}
\vspace{0mm}
\tabletypesize{\footnotesize}
\tablecolumns{5}
\tablewidth{0pt}
\tablecaption{\footnotesize Structural properties of our galactic initial condition \label{tab:1-IC}}
\tablehead{
\colhead{}&\colhead{Dark matter halo}&\colhead{Gas halo (if included in} \vspace{-3mm} &\colhead{Stellar disk}&\colhead{Gas disk}&\colhead{Stellar bulge}\\
\colhead{}&\colhead{}&\colhead{{\scriptsize \sc Gadget-2} / {\scriptsize \sc Gizmo-PSPH})}&\colhead{}&\colhead{}&\colhead{}
}
\startdata
Density profile& \citet{1997ApJ...490..493N} & \citet{1997ApJ...490..493N}  & Exponential & Exponential&\cite{1990ApJ...356..359H} \\
\hline
Structural properties
&$M_{\rm 200, \, crit} =1.074\times10^{12}{\,\rm M_{\odot}}$, $R_{\rm 200}=205.5 \,\,{\rm kpc}$, $c=10$, $v_{\rm 200} = 150\,\, {\rm km\,s^{-1}}$, ${\lambda}=0.04$
&$M_{\rm h, gas} = 3.438\times10^8 {\,\rm M_{\odot}}$ 
&$M_{\rm d,\star}= 3.438\times10^{10}{\,\rm M_{\odot}}$, $r_{\rm d} = 3.43 \,\,{\rm kpc}$, $\,\,z_{\rm d} = 0.1r_{\rm d}$
&$M_{\rm d,gas} = 8.593\times10^{9}{\,\rm M_{\odot}}$, $f_{\,\rm d, gas} =0.2$ 
&$M_{\rm b,\star} = 4.297\times10^{9}{\,\rm M_{\odot}}$, $M_{\,\rm b, \star}/M_{\,\rm d} =0.1$
\\
\hline
No. of particles &10$^{5}$ & $4 \times 10^{3}$ &10$^{5}$ &10$^{5}$ &1.25$\times10^4$\\
\hline 
Particle mass & $m_{\rm DM}=1.254\times10^{7}{\,\rm M_{\odot}}$ & $m_{\rm gas, IC}=8.593\times10^{4}{\,\rm M_{\odot}}$ & $m_{\rm \star, IC}=3.437\times10^{5}{\,\rm M_{\odot}}$ & $m_{\rm gas, IC}=8.593\times10^{4}{\,\rm M_{\odot}}$ & $m_{\rm \star, IC}=3.437\times10^{5}{\,\rm M_{\odot}}$\\
\enddata
\tablecomments{\scriptsize 
The parameters for the gas halo are applicable only for the {\sc Gadget-2} and {\sc Gizmo-PSPH} runs in which the gas halo is included (i.e., those with ``{\tt GasHalo}'' in their run names in Table \ref{tab:2-runs}; see Section \ref{sec:IC2}).  
In mesh-based code {\sc Enzo}, the gas halo is included as a uniform medium around the disk (see Section \ref{sec:IC2}).
All other parameters follow the default disk galaxy initial condition provided by the {\it AGORA} Project \citep{AGORA2016}.  
For more information on the parameters listed above, see Section \ref{sec:initial-condition}.}
\vspace{-3mm}
\end{deluxetable*}

Therefore, in this paper, following up on the {\it AGORA} results, we aim to investigate what causes these inter-code discrepancies in metal distribution and how they can be alleviated. 
We quantitatively compare the metal distribution using various hydrodynamic schemes and analyze several factors that could lead to the discrepancy between mesh-based and particle-based codes --- such as the absence of gas particles in the halo region or the lack of diffusion schemes in the particle-based codes.
We will examine the metal distribution in {\sc Gadget-2} simulations with different hydrodynamic methods and feedback strengths and compare its metal distribution with that of an {\sc Enzo} simulation.
We will also test our proposed prescription to alleviate the inter-code discrepancy in both {\sc Gadget-2} and {\sc Gizmo-PSPH} and discuss its general applicability to particle-based codes. 

This paper is structured as follows. 
In Section \ref{sec:2}, we describe the simulation codes, the initial conditions, and the physics models adopted.
Section \ref{sec:4} compares the metal distributions resulting from different simulation setups (with {\sc Gadget-2} and {\sc Enzo}), focusing on the effect of the initial conditions and the stellar feedback model. 
In Section \ref{sec:5}, we discuss the effect of the metal diffusion scheme in particle-based simulations. 
Then, in Section \ref{sec:6}, we test our proposed prescription to alleviate the inter-code discrepancy in another particle-based code {\sc Gizmo-PSPH} and discuss its general applicability.
Finally, we summarize our findings and discuss future work in Section \ref{sec:7}.

\vspace{1mm}
\section{Numerical Methods}\label{sec:2}
\vspace{1mm}
\subsection{Simulation Codes}  

In this study, we adopt three gravito-hydrodynamics codes widely used in numerical galaxy formation.
We briefly explain the physics and key runtime parameters in each code.

\subsubsection{Mesh-based Code: {\sc Enzo}}

{\sc Enzo} is an Eulerian 3-dimensional structured mesh code with the adaptive mesh refinement (AMR) capability. 
The particle-mesh method is used to compute the gravitational interaction \citep{Hockney&Eastwood1988} while gas dynamics is solved using the 3rd-order accurate piecewise parabolic method \citep[PPM;][]{1984JCoPh..54..174C}.
In this paper, the {\sc Enzo} simulation uses a 64$^{3}$ initial root grid to cover a (1.311 Mpc)$^{3}$ simulation box, achieving an 80 pc spatial resolution with eight additional levels of AMR. 
A cell is refined by a factor of 2 when the mass of the cell is above $m_{\rm gas,IC} = 8.593\times10^{4}\,$M$_{\sun}$ in gas mass, or $8\times m_{\star,\rm IC}$ = $8\times3.437\times10^{5}\,$M$_{\sun}$ in collisionless particle mass. 
Other adopted schemes are largely in line with the recent numerical studies using {\sc Enzo} \citep[e.g.,][]{2019ApJ...887..120K, 2020ApJ...899...25S}. 

\begin{deluxetable*}{lccccc}
\centering
\vspace{0mm}
\tabletypesize{\footnotesize}
\tablecolumns{5}
\tablewidth{0pt}
\tablecaption{List of simulations and key parameters \label{tab:2-runs}}
\tablehead{\colhead{Run name}&\colhead{Simulation code}&\colhead{Gas halo}&\colhead{Stellar feedback $[$ergs per SN$]$ }&\colhead{Diffusion coefficient $C_{\rm d}$ }&\colhead{Stellar mass $[10^9\msun]$}}
\startdata
{\tt Enzo} & {\sc Enzo}&\cmark& 10$^{51}$ &N/A&1.07 \\
{\tt Gad2} & {\sc Gadget-2}&\xmark & 10$^{51}$  &0&1.05 \\
{\tt Gad2-GasHalo} &{\sc Gadget-2} &\cmark & 10$^{51}$  &0&1.06\\
{\tt Gad2-GasHalo+TFB2} & {\sc Gadget-2} &\cmark & $2 \times 10^{51}$  &0&1.00\\
{\tt Gad2-GasHalo+TFB3} &{\sc Gadget-2} &\cmark & $3 \times 10^{51}$  &0&0.97\\
{\tt Gad2-(GasHalo+TFB2)+diff0.3} & {\sc Gadget-2} &\cmark & $2 \times 10^{51}$  &0.006&1.01\\
{\tt Gad2-(GasHalo+TFB2)+diff1} & {\sc Gadget-2} &\cmark & $2 \times 10^{51}$  &0.02&1.11\\
{\tt Gad2-(GasHalo+TFB2)+diff3} & {\sc Gadget-2} &\cmark & $2 \times 10^{51}$  &0.06&1.03\\
{\tt PSPH} & {\sc Gizmo-PSPH}&\xmark & 10$^{51}$ &0&1.02\\
{\tt PSPH-GasHalo+TFB1.8+diff1} & {\sc Gizmo-PSPH}&\cmark &$1.8 \times 10^{51}$ &0.02&0.97\\
\enddata
\tablecomments{\scriptsize List of simulations with different choices of simulation codes, initial conditions (with or without a gas halo; see Section \ref{sec:IC2}), thermal stellar feedback energy, and explicit metal diffusion scheme (with diffusion coefficient $C_{\rm d}$).  
The resulting new stellar mass formed in the first 500 Myr is listed in the rightmost column. 
Unlike particle-based codes, metal diffusion is implicitly performed in the mesh-based code {\sc Enzo}. 
For more information on the items listed here, see Sections \ref{sec:initial-condition} and \ref{sec:2-physics}.  }
\vspace{-3mm}
\end{deluxetable*}

\subsubsection{Particle-based Code: {\sc Gadget-2} and {\sc Gizmo-PSPH}}\label{sec:particle-code}

{\sc Gadget-2} is a tree-particle-mesh SPH code developed by \cite{2005MNRAS.364.1105S}, utilizing a standard density-entropy SPH that manifestly conserves energy, entropy, momentum, and angular momentum.
Yet, the purely density-based SPH scheme may give rise to fictitious pressure on the interface between two media with extreme density contrast \citep{Agertz2007} or damped subsonic turbulence \citep{Bauer&Springel2012}.
Later variants of {\sc Gadget-2} have improved the modelings of complex flows, shocks, or instabilities.
In this study, we test the original {\sc Gadget-2} rather than any specific variants to focus on the fundamental properties of SPH so that our proposed prescription to alleviate the inter-code discrepancy could be applied to later variants. 
We also test another particle-based code, {\sc Gizmo} \citep{2015MNRAS.450...53H}, which includes various hydrodynamic solvers treating the volume components of simulation differently:  density-entropy formalism, pressure-energy formalism, meshless finite mass (MFM), meshless finite volume (MFV), or Eulerian fixed grid schemes. 
For the present study, we experiment with the pressure-energy SPH \citep[hereafter PSPH;][]{Hopkins2013}, which better captures the instability on the surface between different phases than the traditional SPH codes. 
This means that, in the treatment of fluids, the performance of our chosen {\sc Gizmo(-PSPH)} matches that of a contemporary SPH code such as {\sc Gadget-3} widely utilized in the community.

In both {\sc Gadget-2} and {\sc Gizmo-PSPH}, we use the cubic spline kernel \citep{1989ApJS...70..419H} for the softening of the gravitational force with the desired number of neighboring particles $N_{\rm ngb}=32$.
We adopt the Plummer equivalent gravitational softening length $\epsilon_{\rm grav}$ of 80 pc and allow the hydrodynamic smoothing length to reach a minimum of $0.2\epsilon_{\rm grav}$.
In both codes, we include the radiative cooling, star formation and feedback following \citet{AGORA2016}.
We also implement an explicit metal diffusion scheme following \cite{2018MNRAS.480..800H} in the public version of {\sc Gadget-2}.
For detailed explanation on these and other baryon physics included, we refer the readers to Section \ref{sec:2-physics}.  

\subsection{Initial Condition}\label{sec:initial-condition}

We have adopted the disk initial condition (IC) provided by the {\it AGORA} Project \citep{AGORA2016} that models an idealized Milky Way-mass disk galaxy of $M_{\rm 200, \,crit} =1.074\times10^{12}{\,\rm M_{\odot}}$ in isolation.
The IC includes a dark matter halo that follows the Navarro-Frenk-White profile \citep[NFW;][]{1997ApJ...490..493N}, an exponential disk of stars and gas, and a stellar bulge following the Hernquist profile \citep{1990ApJ...356..359H}.\footnote{The {\it AGORA} initial conditions are publicly available at \url{http://sites.google.com/site/santacruzcomparisonproject/blogs/quicklinks/}.}
The detailed structural parameters of the IC are listed in Table \ref{tab:1-IC}.
In what follows, we discuss, in particular, the gas distribution in the original {\it AGORA} IC and our modified IC.   

\subsubsection{Gas Distribution In The Original {\it AGORA} IC}\label{sec:IC}

For the mesh-based code {\sc Enzo}, the gas density field of the disk is initialized with an exact analytic formula 
\begin{equation}
\rho_{\rm d,gas}(r,z) =  \rho_{0}\,e^{-\frac{r}{r_{\rm d}}}\cdot e^{-\frac{\abs{z}}{z_{\rm d}}}  \label{eq:disk}
\end{equation}
where $r$ is the cylindrical radius, $z$ is the vertical distance from the disk plane, $r_{\rm d} = $ 3.432 kpc, $z_{\rm d} = 0.1\,r_{\rm d}$, and $\rho_{0}=M_{\rm d, \,gas}/(4 \pi r_{\rm d}^{2}\,z_{\rm d}$) with $M_{\rm d,\,gas} = 8.593\times 10^{9}\msun$. 
On the other hand, for the particle-based codes {\sc Gadget-2} and {\sc Gizmo-PSPH}, the {\it AGORA} IC provides a text file of the initial positions of gas particles in the disk that follow Eq.(\ref{eq:disk}). 
Notably, the gas particles are absent in the halo region.
For both mesh-based and particle-based codes, the initial temperature in the disk is set to $10^{4}\,{\rm K}$ and the initial metallicity to $Z_{\rm disk} = 0.02041$. 
Finally, for the mesh-based code only, the gas density in the halo is set to a constant $n_{\rm H}=10^{-6}\,{\rm cm}^{-3}$ to avoid a zero value in the cells, with an initial metallicity of $Z_{\rm halo}=10^{-6}\,Z_{\rm disk}$ and temperature of 10$^{6}$ K. 

\subsubsection{Gas Halo In The Modified {\it AGORA} IC}\label{sec:IC2}

In the original {\it AGORA} IC employed in \citet{AGORA2016} to model an idealized Milky Way-mass disk galaxy, a gas halo is included only in mesh-based codes, albeit with a negligible density $n_{\rm H}=10^{-6}{\rm cm}^{-3}$ as mentioned in the previous section.
In this paper, we demonstrate that this small difference can cause substantial discrepancies in the baryonic properties of a galaxy during the 500 Myr evolution, especially in the halo. 
In this section, we will first argue that a sufficiently-resolved gas halo is necessary for particle-based simulations to properly model the metal transport and mixing in the circumgalactic medium (CGM), and explain the modified IC adopted in some of our simulations. 

Consider a situation in which the disk gas is being pushed by strong SN winds and is on the verge of leaving the disk's ISM into the halo. 
Without any gas particles in the halo \citep[e.g., in the particle-based simulations described in][]{AGORA2016}, the supersonic gas outflows do not experience any pressure that impedes its motion.  
Hence the outflow continues to move into the vacuum while losing little or no momentum.
On the contrary, in simulations with a gas halo, the outflow will be slowed down or sometimes even severely suppressed.
As we will demonstrate in later sections, simulators or those who analyze simulations should be cautioned that this discrepancy may cause substantial deviations in the baryonic properties in the ISM and CGM.
In addition, without the gas halo, the metals in disk gas particles have no particle to diffuse into in the halo region.  
Therefore, until the halo is populated with (a few) gas particles expelled from the disk by SN winds, the halo rarely becomes metal-enriched.  

\begin{figure}
\vspace{0mm}
\centering
\includegraphics[width=0.42\textwidth]{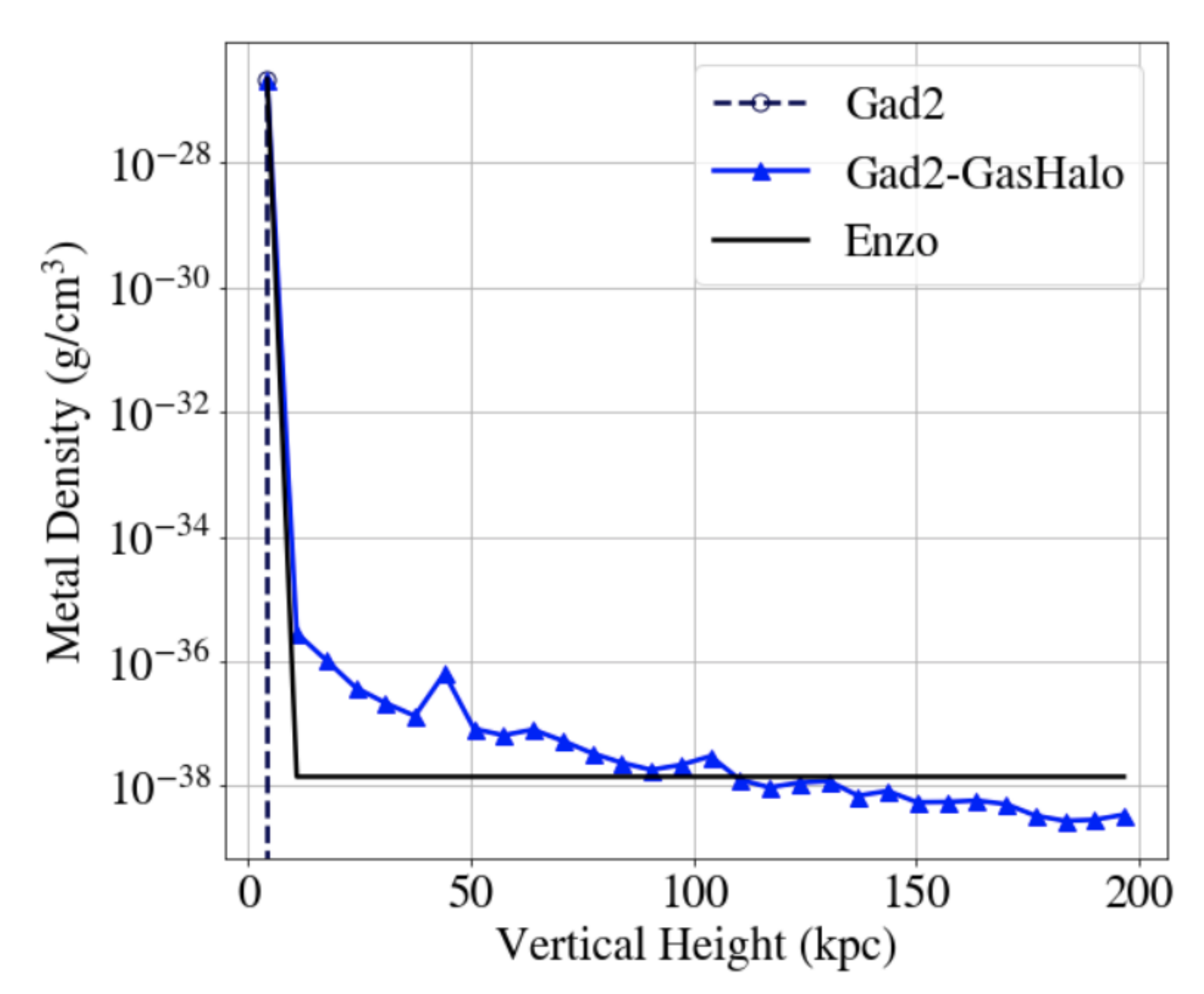}
\vspace{-1mm}
\caption{The 0 Myr profiles of density-weighted metal density as functions of the vertical distance from the disk plane in the {\tt Gad2}, {\tt Gad2-GasHalo}, and {\tt Enzo} runs (see Table \ref{tab:2-runs}).
The {\tt Gad2} run does not contain any gas in the halo region while the initial gas distribution in {\tt Gad2-GasHalo} is approximately comparable with that in {\tt Enzo}.
See Section \ref{sec:IC2} for more information. 
}
\vspace{0mm}
\label{fig01-ic}
\end{figure}

For these reasons, in this study, we test a different IC to model an isolated galaxy with the particle-based codes {\sc Gadget-2} and {\sc Gizmo-PSPH}, that allocates additional gas particles in the halo region.  
Gas particles are placed in the halo following the NFW profile in a way that they approximately match the initial halo gas density in the mesh-based code's IC,  $n_{\rm H}=10^{-6}\,{\rm cm}^{-3}$.
We match the mass of an individual gas particle in the halo to that in the disk, $m_{\rm gas, IC}=8.593\times10^{4}\msun$, resulting in a total of 4,000 gas particles in the halo (see Table \ref{tab:1-IC}).  
The initial metallicity is set to $Z_{\rm halo}=10^{-6}\,Z_{\rm disk}$ and the temperature to 10$^{6}$ K. 
The particle-based simulations that utilize this revised IC are denoted with ``{\tt GasHalo}'' in their run names in Table \ref{tab:2-runs}.  
The initial metal distributions of the original {\it AGORA} IC and the revised IC are displayed in Figure \ref{fig01-ic}, along with that for the mesh-based codes.
In the {\tt Gad2} run's IC \citep[identical to the particle-based codes' ICs in][]{AGORA2016} the halo is free of gas (vertical height $z > 5$ kpc), while the {\tt Gad2-GasHalo} and {\tt Enzo} runs feature nonzero gas density and metals in the halo region.

\subsection{Baryon Physics} \label{sec:2-physics}

We consider all the baryon physics that are relevant in the process of galaxy formation by closely following the previous {\it AGORA} disk comparison \citep{AGORA2016}, along with an optional scheme for explicit metal diffusion.

\subsubsection{Cooling, UV Background, Jeans Pressure Support}\label{sec:pressuresupport}

The radiative cooling and heating rates for the gas are calculated with the {\sc Grackle} library \citep{2017MNRAS.466.2217S}. 
We adopt {\sc Grackle}'s ionization equilibrium mode with the \citet{HaardtMadau12} UV background radiation at $z=0$  --- i.e., the gas cooling rate is determined by its density, temperature and metallicity in the ionization levels satisfying the equilibrium state  {using {\sc Cloudy}} \citep{Ferland2013}.
In addition, we include the Jeans pressure floor to avoid any artificial collapse and numerical fragmentation  \citep{1997ApJ...489L.179T}. 
The Jeans pressure is determined as
\begin{equation}
    P_{\rm Jeans} = \frac{G}{\gamma\pi}N^{2}_{\rm Jeans}\rho^{2}_{\rm gas}\Delta x^{2},
\end{equation}
where $\gamma=5/3$ is the adiabatic index, $G$ is the gravitational constant, and $\rho_{\rm gas}$ is the gas density. 
Here, $\Delta x$ is equivalent to the spatial resolution (or its proxy) carried by each simulation code ---  that is, the finest cell size in {\sc Enzo}, the smoothing length $h_{s\rm ml}$ in {\sc Gadget-2}, and the radius of the ``effective volume'' of a cell in {\sc Gizmo-PSPH}, $(4\pi/(3 N_{\rm ngb}))^{1/3}h_{\rm sml}$.
Correspondingly, we set the controlling parameter $N_{\rm Jeans}$ to 4.0, 4.2 and 6.3 for {\sc Enzo}, {\sc Gadget-2} and {\sc Gizmo-PSPH}, respectively, to provide a similar amount of pressure across the codes.
These choices of $N_{\rm Jeans}$ are in line with \citet[][see their Section 3.1]{AGORA2016}, and lead to model galaxies producing similar stellar masses of $\sim 10^9 \msun$ in the first 500 Myr \citep[see Table \ref{tab:2-runs}; see also Figure 26 of][]{AGORA2016}.

\subsubsection{Star Formation and Feedback}\label{sec:sf}

Gas parcels that are denser than a threshold, $n_{\rm H}=10\,{\rm cm}^{-3}$, spawn stars at a rate following the local Schmidt law
\begin{equation}
        \dv{\rho_{\star}}{t} = \frac{\epsilon_{\star} \rho_{\rm gas}}{t_{\rm ff}},
\end{equation}
where $\rho_{\star}$ is the stellar density, $t_{\rm ff} = (3\pi/(32\, G\rho_{\rm gas}))^{1/2}$ is the local free-fall time, and $\epsilon_{\star} = 0.01$ is the star formation efficiency per free-fall time.
5 Myr after their formation, star particles inject thermal energy, mass, and metals into their surrounding ISM in an attempt to describe Type II SN explosions.   
Following \citet{Chabrier2003} initial mass function (IMF), we assume that for stars with a mass range of 8 - 40 $\msun$, a single SN event occurs per er{every} 91 $\msun$ of stellar mass formed, releasing 2.63 $\msun$ of metals and 14.8 $\msun$ of gas (including metals).
To probe the difference in the efficiency of stellar feedback between mesh-based and particle-based codes --- especially in the context of metal transport ---  we test various thermal energy values of the stellar feedback: 10$^{51}$, 2$\times10^{51}$, and 3$\times10^{51}$ ergs per SN, labelled as the {\tt Gad2-GasHalo/+TFB2/+TFB3} run (see Table \ref{tab:2-runs}). 

\begin{figure*}
\centering
\vspace{-1mm}
\includegraphics[width=14.4cm]{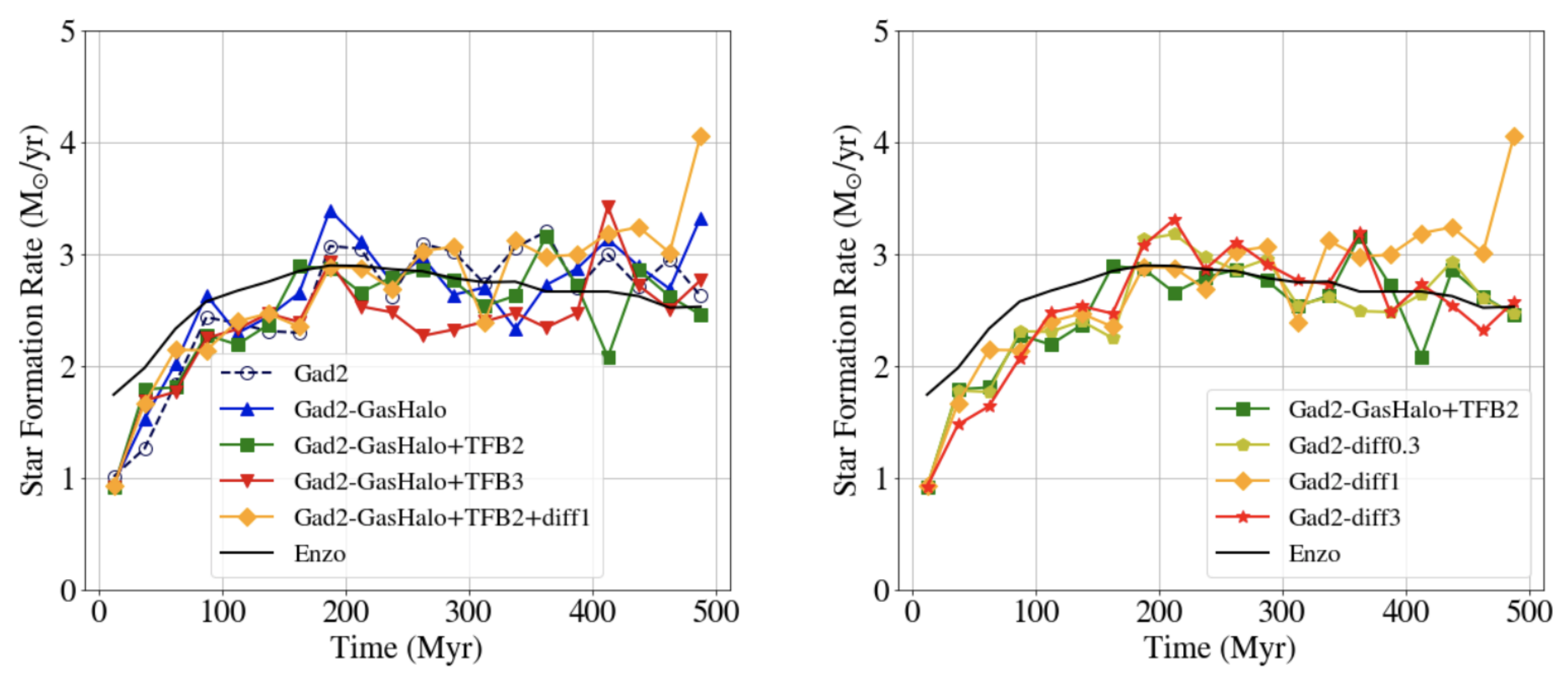}
\vspace{-2mm}
\caption{
The star formation rates (SFRs) for the first 500 Myr with different simulation setups (different codes, ICs, feedback strengths, with or without metal diffusion, etc.; {\it left panel}) and different metal diffusion coefficients ({\it right panel}).  
On the right panel, all {\sc Gadget-2} runs are with the gas halo and the thermal feedback energy of $2\times 10^{51}$ ergs per SN (i.e., {\tt GasHalo+TFB2}). 
Most of the runs exhibit similar SFRs throughout the simulation. 
See Table \ref{tab:2-runs} for the list of our simulations, and Section \ref{sec:3} for more information on this figure.}
\label{fig02-sfr}
\vspace{3mm}
\end{figure*}

\subsubsection{Explicit Metal Diffusion In Particle-based Simulations}\label{sec:diffusion}

In particle-based simulations, once a metal field is assigned to a particle in the IC, its value never changes unless the particle is directly affected by a SN bubble.
This means that a naive particle-based approach does not capture the inter-particle diffusion of metals.  
Evidently, many particle-based code groups studying metal transport in a galaxy-scale simulation have devised a way to model how the metal is mixed.  
The diffusion scheme has also been shown indispensable to match the observed scatters of metal element abundances, such as alpha and r-process elements \citep[see, e.g.,][]{Revaz2016, Escala2018, Dvorkin2020}. 
In this study, we consider an explicit turbulent metal diffusion scheme in {\sc Gadget-2} and {\sc Gizmo-PSPH}, and compare the metal distribution in the galactic disk and halo, with and without the scheme. 

We adopt the metal diffusion scheme used in \citet{2018MNRAS.480..800H} and  \citet{Escala2018}, which itself is based the Smagorinsky-Lilly model \citep{Smagorinsky1963, Shen2010}.
In brief, the model estimates the subgrid diffusion effect driven by the velocity shear between particles, assuming that the local diffusivity is dependent on the velocity shear and the resolution scale.
That is, the metal diffusion between the particles in a shear motion with respect to each other is
\begin{align}
    \pdv{M_{i}}{t} &+ \div{(D \nabla M_{i})}=0,  \\
    D &= C_{\rm d}\,||\boldsymbol{S}_{ ij}||\,h^{2},
\end{align}
where $M_{i}$ is the scalar field (metallicity) of the $i$-th particle, $h$ is the effective measurement scale (which we choose to set to the SPH kernel size $h_{s\rm ml}$), $||\cdot ||$ is the Frobenius norm, and $C_{\rm d}$ is the diffusion coefficient that is proportional to the Smagorinsky-Lilly constant, calibrated by numerical simulations based on the Kolmogorov theory. 
And the symmetric trace-free tensor $S_{ij}$ is given by
\begin{align}
    \boldsymbol{S}_{ij} = \frac{1}{2}\{\frac{\partial v_j}{\partial x_i}\, +\frac{\partial v_i}{\partial x_j}\}- \frac{1}{3}\delta_{ij}{\rm Tr} (\frac{\partial v_i}{\partial x_j})
\end{align}
where $i, j = \{x, y, z\}$, $v_{i}$ is the velocity vector for each gas particle, and, $x_i$ is the spatial coordinate.
Thus, the turbulent diffusion becomes efficient in large eddies where the shear drives local fluid instabilities. 
We caution the readers that this simplistic diffusion model is always dissipative, and the backscattering from small to large scales is not possible.
The single controlling parameter in this model, the diffusion coefficient $C_{\rm d}$, has difficulty in properly describing various types of turbulent flows, either.
For example, the model may overestimate the diffusion in the laminar flows \citep{Rennehan2019}.
Additionally, \cite{Colbrook2017} showed that the turbulent diffusivity is dependent on the scale of eddies.

Despite these limitations, we adopt the Smagorinsky-Lilly model to show that including such a simple diffusion model can mitigate the known discrepancy between mesh-based and particle-based codes.
Different authors have used different values for $C_{\rm d}$, ranging from 0.003 in \citet{Escala2018} to 0.05 in \citet{Shen2010}.\footnote{Readers should note that \citet{Escala2018} studied isolated dwarf galaxies in cosmological zoom-in simulations, while \citet{Shen2010} examined intergalactic-scale phenomena in larger-box simulations.  This may explain the wide range of diffusion coefficients chosen by different authors.}
Given such a wide range of values found in the literature, here we test $C_{\rm d}=$ 0.006, 0.02, and 0.06 which are labelled as the {\tt Gad2-diff0.3}, {\tt Gad2-diff1}, and {\tt Gad2-diff3} run, respectively (see Table \ref{tab:2-runs}).

\vspace{2mm}

\section{Metal Distribution In Halo:  Dependence on Initial Gas Distribution And Feedback Strength}\label{sec:4}

\vspace{1mm}

Using a suite of simulations listed in Table \ref{tab:2-runs}, we compare how the spatial distribution of metals differs between different types of simulations --- first focusing on how the existence of a halo gas and the different feedback strengths change the extent of transported metals in particle-based simulations.  

\begin{figure*}
\vspace{1mm}
\centering
\includegraphics[width=18.5cm]{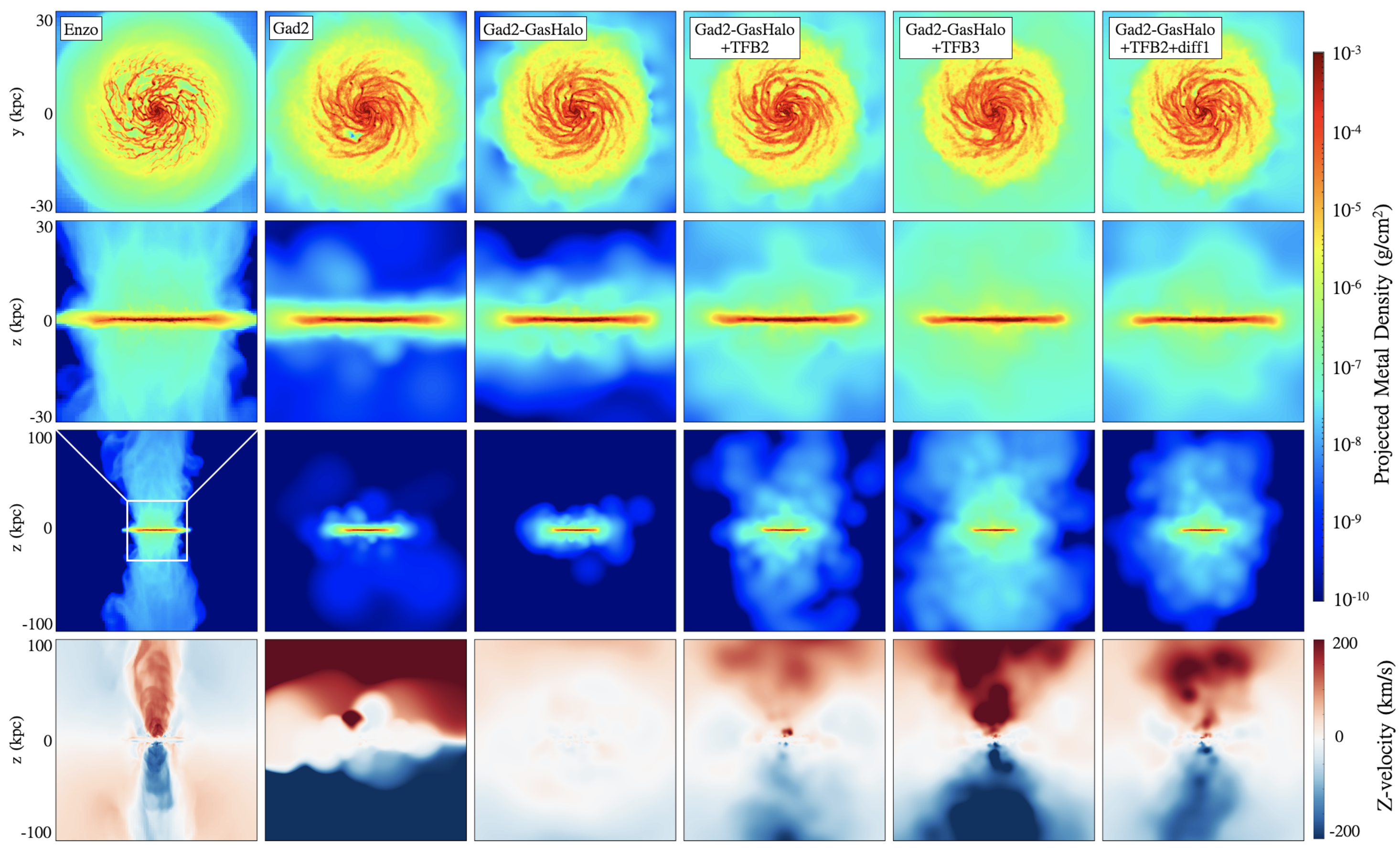}
\vspace{-5mm}
\caption{
$500$ Myr snapshots of our isolated disk simulations using different ICs, stellar feedback, and diffusion schemes. 
Face-on ({\it 1st row}), edge-on ({\it 2nd row}) and wider edge-on ({\it 3rd row}) projections of metal density, and density-weighted edge-on projection of the vertical velocity ({\it 4th row}).
The {\sc Gadget-2} simulation without the halo gas --- the {\tt Gad2} run (see Table \ref{tab:2-runs}); the same runtime condition as the particle-code runs in \citet{AGORA2016} --- substantially differs from the mesh-based {\tt Enzo} run in the metal distribution in the halo.  
In contrast, another {\sc Gadget-2} simulation, but this time with a gas halo, more feedback energy, and explicit metal diffusion scheme --- the {\tt Gad2-GasHalo+TFB2+diff1} run  --- is  comparable with the {\tt Enzo} run.  
See Table \ref{tab:2-runs} for the list of our simulations, and Section \ref{sec:projection-fb} for more information on this figure.}
\label{fig03-proj}
\vspace{3mm}
\end{figure*}

\subsection{Comparison of Star Formation Rates}\label{sec:3}

Metals in galaxies are produced by stars. 
Therefore, in order to compare the spatial distribution of metals between different simulations, it is necessary to establish a baseline in which all simulations exhibit similar star formation histories in the timespan considered. 
In Figure \ref{fig02-sfr}, we show the SFRs in simulations of different hydrodynamic solvers, different ICs, different feedback strengths, and different diffusion coefficients.
As noted in Section \ref{sec:pressuresupport}, the Jeans pressure support for each code ({\sc Enzo}, {\sc Gadget-2}, {\sc Gizmo-PSPH}) are set in such a way that the runs show similar SFRs and the value $N_{\rm Jeans}$ is in line with the previous {\it AGORA} comparison \citep{AGORA2016}.  
As a result, despite the differences in the simulation setup, most of the runs analyzed in this article exhibit similar star formation histories within a few tens of percents at all times, totaling a stellar mass of $\sim 10^9\msun$ in the first 500 Myrs (see the rightmost column of Table \ref{tab:2-runs}).\footnote{We hereafter compare the suite of simulations at 500 Myr, which is a timescale sufficient to observe the turbulent diffusion effect \citep[see, e.g., Eqs.(10)-(11) in][]{Williamson2016}, but is also motivated by the previous {\it AGORA} comparison \citep{AGORA2016}.}

\begin{figure*}[ht!]
\vspace{0mm}
\centering
\includegraphics[width=18.3cm]{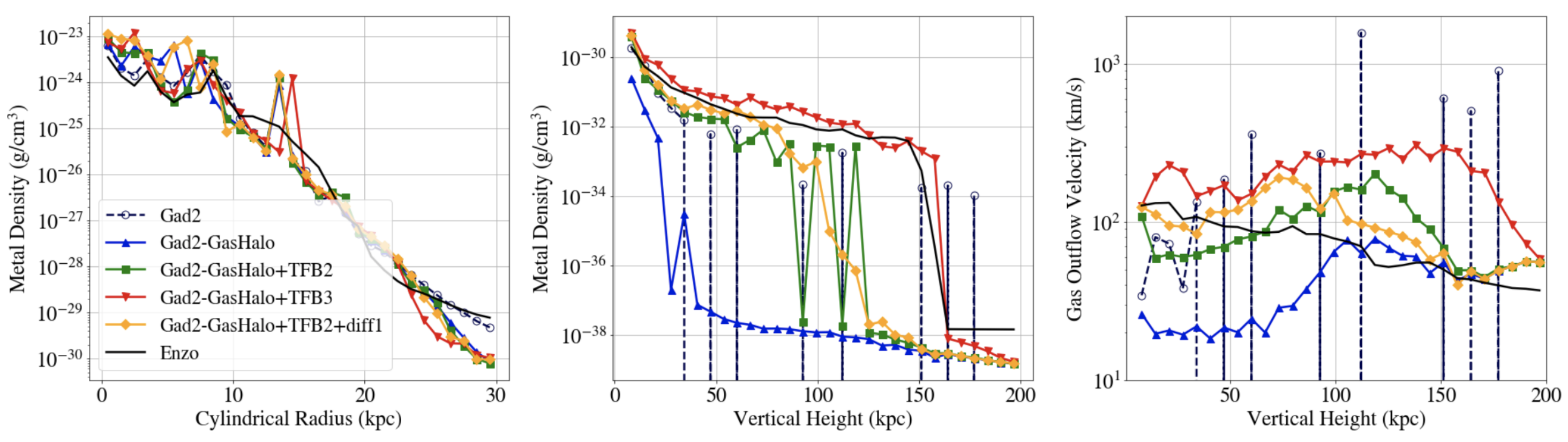}
\vspace{-6mm}
\caption{
Density-weighted metal density profiles as functions of the disk's cylindrical radius ({\it left panel}) and the vertical height from the disk plane ({\it middle panel}), for different simulation setups at 500 Myr.
Also shown are the gas outflow velocity profiles as functions of the vertical height ({\it right panel}).
Here, the cylindrical profile is made for the disk gas only, defined as the region where the vertical distance from the disk plane $z$ is less than $5\,{\rm kpc}$.
Most of the runs exhibit similar profiles in the cylindrical radial direction, while the vertical distribution of metals is highly dependent on the simulation setups. 
See Section \ref{sec:profile-fb} for more information on this figure.
}
\label{fig04-prof-1}
\vspace{4mm}
\end{figure*}

Note that even the seemingly important differences such as in the stellar feedback strength or in the diffusion coefficient introduce only marginal changes in SFRs in Figure \ref{fig02-sfr}.
For example, the run with higher thermal energy suppresses star formation only slightly when compared with the one with lower energy (compare the {\tt Gad2-GasHalo/+TFB2/+TFB3} runs in Table \ref{tab:2-runs} and Figure \ref{fig02-sfr}).  
Meanwhile, the coefficient of metal diffusion does not substantially affect the produced stellar mass  (compare the {\tt Gad2-(GasHalo+TFB2)+diff0.3/diff1/diff3} runs in Table \ref{tab:2-runs}).
Since all the simulation produce a similar amount of stars --- and thus metals --- we can now conjecture that any difference in the spatial distribution of metals is due to the difference in how each simulation transports metals in and out of the galactic disk (e.g., different feedback strength, inherent differences in hydrodynamics), not because any one simulation harbors a larger/smaller amount of metals.  

\subsection{Overview: Metal Distribution and Outflow Velocity}\label{sec:projection-fb}

\begin{figure*}
\vspace{0mm}
\centering
\includegraphics[width=15cm]{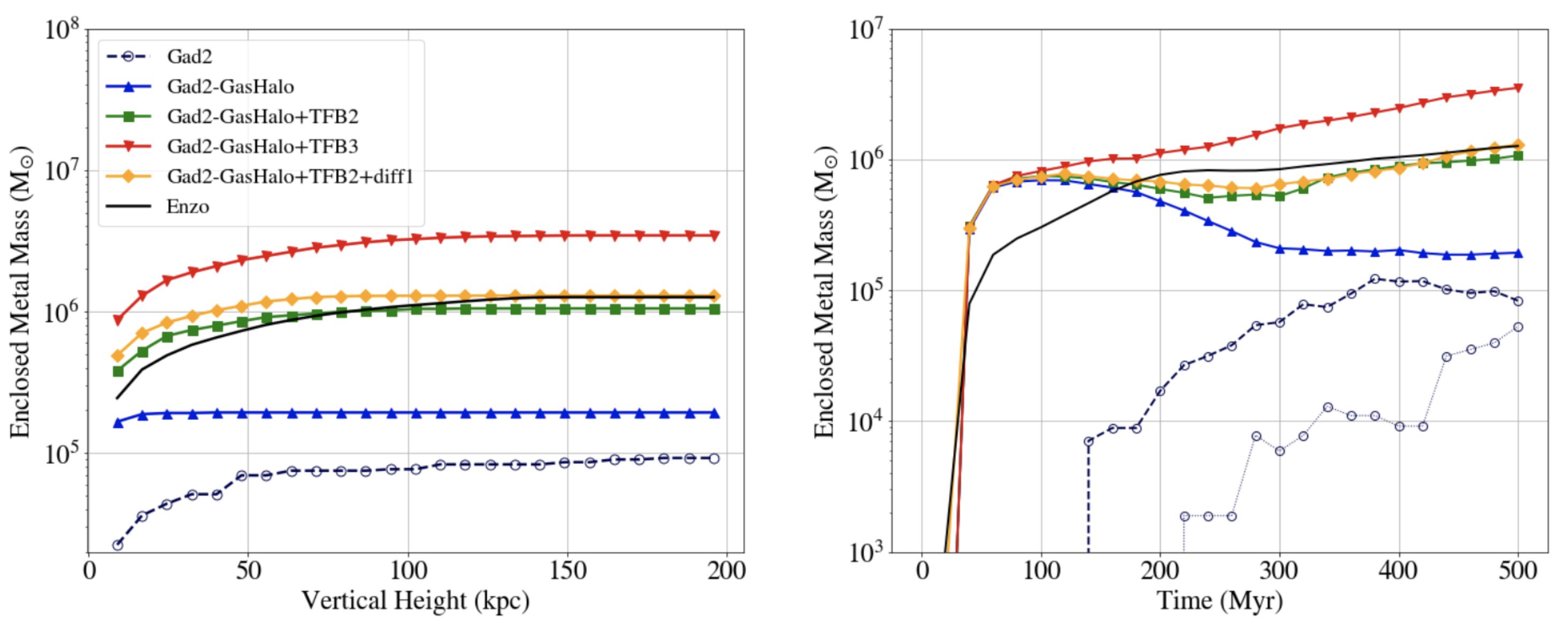}
\vspace{-2mm}
\caption{
Enclosed metal mass profiles as functions of the vertical height from the disk plane  for different simulation setups at 500 Myr ({\it left panel}).  
Also shown is the time evolution of total metal mass in the halo ({\it right panel}).
Here, the halo is defined as the region where the vertical distance from the disk plane $ z > 5\,{\rm kpc}$ or the radial distance from the disk center $r > 60 \,{\rm kpc}$.
The thin dotted line denotes the enclosed metal for the {\tt Gad2} run in the region of $R_{200} < r < 2R_{200}$ where $R_{200} = 205.5$ kpc.
See Section \ref{sec:profile-fb} for more information on this figure.
}
\label{fig04-prof-2}
\end{figure*}

Figure \ref{fig03-proj} displays the face-on and edge-on projections of metal density and the vertical velocity of gas outflows from the disk plane at 500 Myr after the simulation starts. 
In the face-on view, all simulations show a similar distribution of metals along the spiral arms.
The edge-on view of metal distribution, however, varies substantially depending on the simulation setup.
The first stark contrast is between the {\tt Enzo} run and the {\tt Gad2} run (1st and 2nd column in Figure \ref{fig03-proj}, respectively).
As previously reported by the {\it AGORA} Collaboration \citep{AGORA2016}, in a  particle-based simulation with neither the gas halo nor an explicit diffusion scheme, metals are inevitably scarce in the halo away from the disk  (i.e., 2nd and 3rd rows of the {\tt Gad2} run).  
This is because the halo is only populated with very few (metal-enriched) gas particles ejected from the disk --- as discussed in Section \ref{sec:IC2} and will become more obvious in later sections.  
The enclosed metal mass in the halo region (vertical distance from the disk $z > 5$ kpc) is about $\gtrsim$ 30 times lower in the  {\tt Gad2} run than in the {\tt Enzo} run (see also the left panel of Figure \ref{fig04-prof-2}; to be discussed in details in Section \ref{sec:profile-fb}).
Despite having a very metal-poor halo, the {\tt Gad2} run shows the fastest gas outflows from the disk among all the runs, in the bottom row of Figure \ref{fig03-proj} that displays the density-weighted projection of vertical velocity. 
This counter-intuitive result is due to the unphysical nature of the {\tt Gad2} run's IC, and to the fact that the halo region contains only a few gas particles with an extremely high velocity (see also the right panel of Figure \ref{fig04-prof-1}; to be further discussed in details in Section \ref{sec:profile-fb}). 
In the absence of gas in the halo region, the high-velocity SN ejecta travels into the halo without experiencing any pressure that impedes its motion.  
Not suffering any deceleration, these high-velocity gas particles may reach hundreds of kiloparsecs away from the disk.

Therefore, to rectify the unphysical results of the {\tt Gad2} run, another IC has been tested, which now includes the gas halo around the disk --- i.e., the {\tt Gad2-GasHalo} run (3rd column in Figure \ref{fig03-proj}; see also Section \ref{sec:IC2}).
As we compare the {\tt Gad2} and the {\tt Gad2-GasHalo} run, we first find that in terms of metals in the halo, the {\tt Gad2-GasHalo} run is hardly different from the {\tt Gad2} run (3rd row).
This is because the additional halo gas particles still cannot receive metals unless there is an explicit way for the metals to diffuse into the halo.
It is also because the SN ejecta cannot easily penetrate the gas halo as it did in the {\tt Gad2} run.  
The halo gas applies ram pressure on the gas outflows at the disk-halo boundary and restricts the reach of the metal-enriched ejecta.\footnote{As noted in Section \ref{sec:particle-code}, the original {\sc Gadget-2} tends to suppress fluid instabilities due to the shielding effect between the two media with extreme contrast in density \citep{Agertz2007}.  Later variants of {\sc Gadget-2} have improved to capture such instabilities, which can affect the penetration of high-velocity winds into the halo \citep{Hopkins2013}.}
As a result, the galactic outflow becomes very weak in the {\tt Gad2-GasHalo} run (bottom row of Figure \ref{fig03-proj}), and the metal-enriched ejecta remains near the galactic disk (3rd row).  

\begin{figure*}
\vspace{0mm}
\centering
\includegraphics[width=15.2cm]{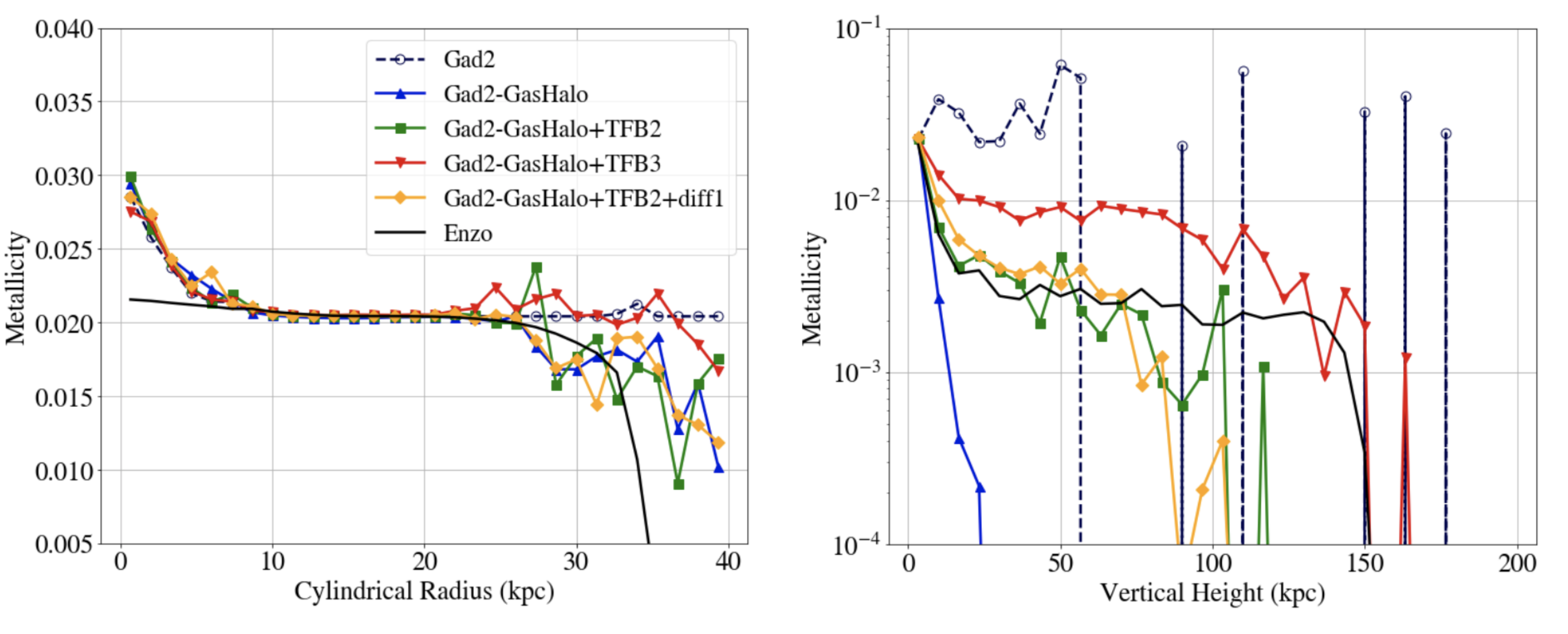}
\vspace{-3mm}
\caption{
Mass-weighted metallicity profiles as functions of the disk's cylindrical radius ({\it left panel}) and the vertical height from the disk plane ({\it right panel}), for different simulation setups at 500 Myr.  
Here, the cylindrical profile is made for the disk gas only, defined as the region where the vertical distance from the disk plane $z$ is less than $5\,{\rm kpc}$.
The shapes of the metallicity profiles are affected by the existence of the halo gas and by the amount of stellar feedback energy.
See Section \ref{sec:profile-fb} for more information on this figure.
}
\label{fig04-prof-3}
\vspace{2mm}
\end{figure*}

\begin{figure*}[ht!]
\vspace{-2mm}
\centering
\includegraphics[width=18.2cm]{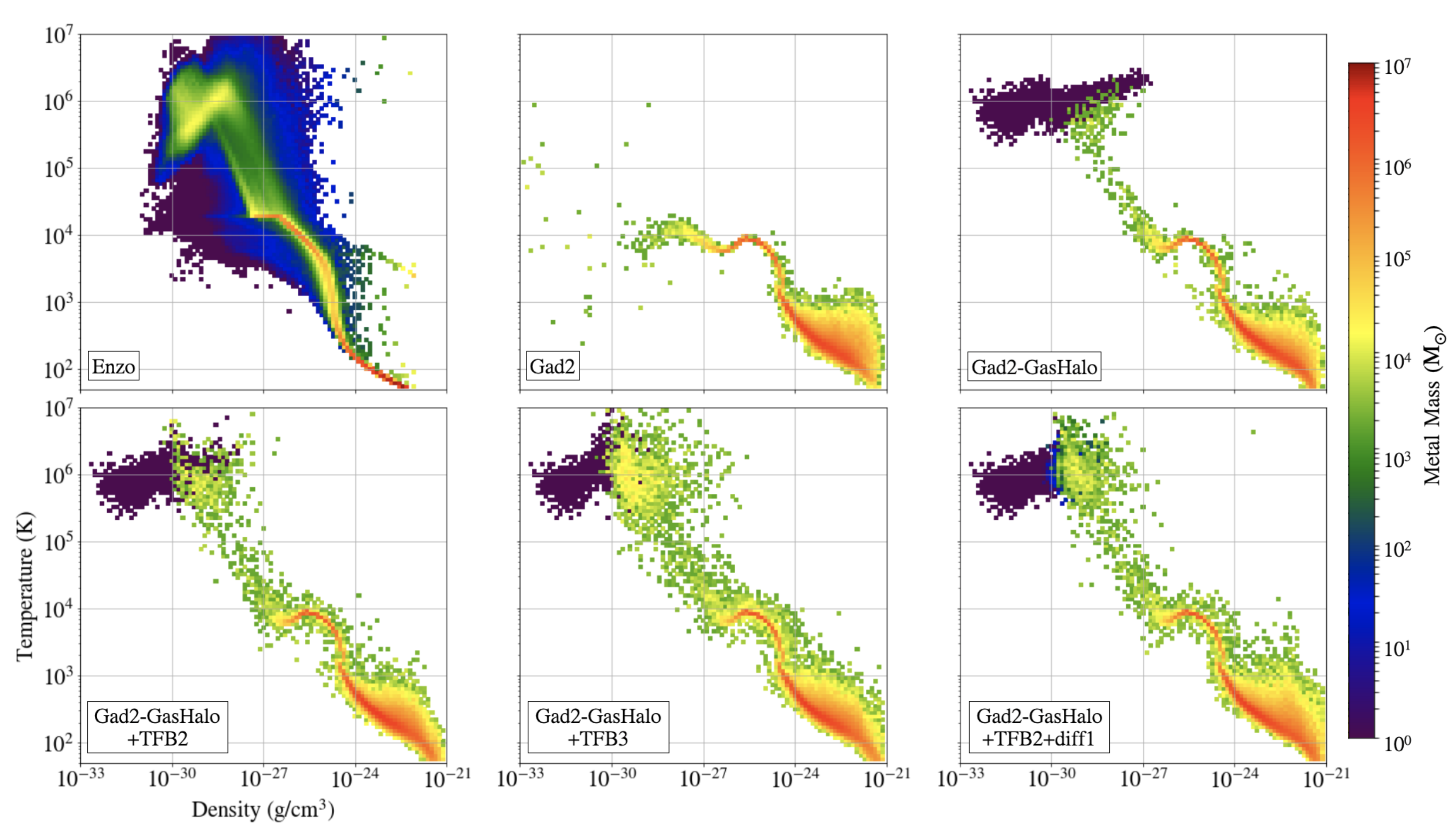}
\vspace{-6mm}
\caption{
Two-dimensional probability distribution functions (PDFs) of metal mass on the density-temperature plane for different simulation setups at 500 Myr. 
The color represents the mass of the metals in each two-dimensional bin. 
Only with the gas halo sufficiently resolved in the IC (i.e., {\tt Gad2-GasHalo/+TFB2/+TFB3/+TFB2+diff1} runs) can the particle-based simulations identify  the hot diffuse gas surrounding the disk at $[\sim 10^{-29} \,{\rm g\,cm^{-3}}, \,\sim10^6 \,{\rm K}]$.
The metal mass distribution in the PDF, especially in the halo, heavily depends on the thermal stellar feedback energy as well. 
See Table \ref{tab:2-runs} for more information on our list of simulations, and Section \ref{sec:pdf-fb} for more information on this figure.
}
\label{fig05-pdf}
\vspace{3mm}
\end{figure*}

In other words, our experiment suggests that particle-based codes may require more stellar feedback energy than mesh-based codes to launch galactic outflows into the gas halo.  
Indeed, the {\sc Gadget-2} simulation with twice the thermal feedback energy --- i.e., the {\tt Gad2-GasHalo+TFB2} run (4th column in Figure \ref{fig03-proj}) --- shows a similar metal distribution and outflow velocity to the {\tt Enzo} run.\footnote{Although the outflow is with a larger opening angle than in the {\tt Enzo} run.} 
Comparing the runs with varying thermal feedback energies --- {\tt Gad2-GasHalo/+TFB2/+TFB3} runs --- we find that the metal enrichment in the halo is highly sensitive to the feedback strength.
The amount of thermal energy injected directly determines the momentum of SN ejecta, and consequently, the mass of metal-enriched gas in more turbulent bubbles.
The increased turbulence enables more metal-enriched gas to be coupled with large momentum, allowing the SN ejecta to escape from the disk easily.
In particle-based simulations, this process may require more energy than in mesh-based ones, due to the inherent inter-code discrepancies in how the thermal feedback energy is distributed in the neighborhood of newly-born stars, and how the Riemann problem is solved at the disk-halo boundary.
For fluids in vastly different phase --- e.g., SN hot bubbles in the cold-dense gas clouds --- in particle-based simulations, the density of the dilute fluid can be overestimated by the SPH kernels in insufficient resolution, which gives rise to overcooling and inhibiting the development of hot gas \citep{Marri2003,Creasey2011}.\footnote{Note that we have only tested the thermal feedback prescription based on \citet{AGORA2016} (see Section \ref{sec:sf} for the details).  Many particle-based code simulations try different strategies to model the stellar feedback, such as kinetic feedback, stochastic feedback, radiation from young stars, delayed cooling,  etc. \citep[e.g.,][]{Dalla2012, revaz_pushing_2018, 2018MNRAS.480..800H, 2019MNRAS.484.2632S}. Different feedback strategies may help deposit the energy into the ISM more efficiently.}   
 
Lastly, we include the explicit metal diffusion scheme (see Section \ref{sec:diffusion}) to allow metals of highly-enriched gas particles to slowly diffuse into the less-enriched neighbors.
Comparing the {\tt Gad2-GasHalo+TFB2} and {\tt Gad2-GasHalo+TFB2+diff1} run in Figure \ref{fig03-proj}, we discover that the metal distribution in space does not highly depend  on the diffusion scheme.
However, in Section \ref{sec:MDF-diff} we will demonstrate why the diffusion scheme must be included.

\subsection{Spatial Profiles of Metals and Its Evolution In Time} \label{sec:profile-fb}

Thus far, using Figure  \ref{fig03-proj} we have shown that the presence of the gas particles in the halo region, despite its negligible density, affects the metal distribution therein.  
We have also demonstrated that the amount of metals expelled from the disk depends on the stellar feedback energy.
In this subsection, we further investigate these points quantitatively.

\begin{figure*}[ht!]
\centering
\includegraphics[width=15cm]{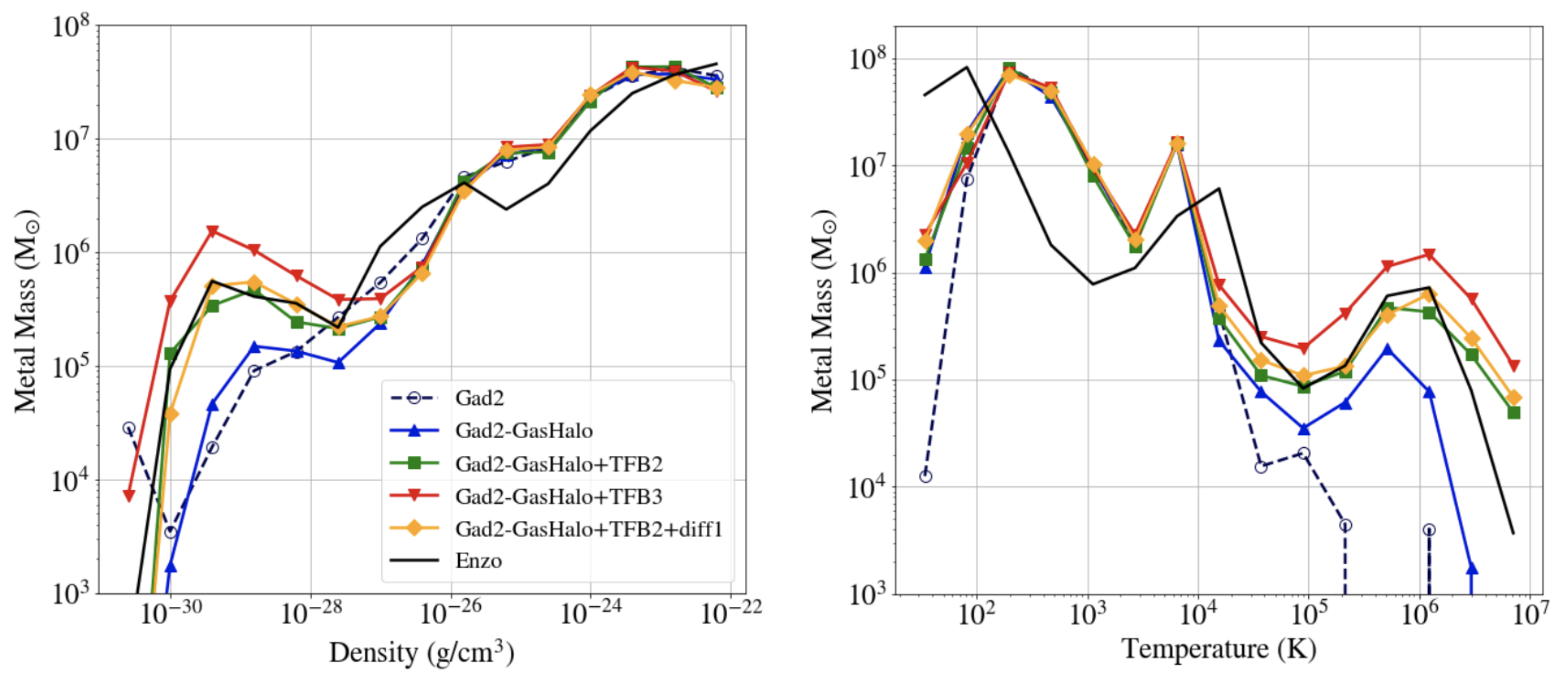}
\vspace{-2mm}
\caption{
The probability distribution functions (PDFs) of metal mass in gas density ({\it left panel}) and temperature ({\it right panel}), for different simulation setups at 500 Myr. 
Only with the gas halo sufficiently resolved in the IC (i.e., {\tt Gad2-GasHalo/+TFB2/+TFB3/+TFB2+diff1} runs) can the particle-based simulations identify the hot diffuse gas in the halo at $10^{-30} - 10^{-27} \,{\rm g\,cm^{-3}}$ and  $10^{4} - 10^{7} \,{\rm K}$.
The amount of metal mass in this hot diffuse medium is dictated by the strength of stellar feedback. 
See Section \ref{sec:pdf-fb} for more information on this figure.
}
\label{fig06-1d-pdf}
\vspace{3mm}
\end{figure*}

Figure \ref{fig04-prof-1} illustrates the density-weighted metal density profiles and the velocity of gas outflows perpendicular to the disk plane 500 Myrs after the simulation starts. 
As observed in Figure \ref{fig03-proj}, the metal density profiles in the disk's radial direction (left panel of Figure \ref{fig04-prof-1}) are similar across different simulation setups. 
In contrast, the metal density distribution in the halo and the amount of metal transported to the halo are notably different between the runs.
The middle and right panels of Figure \ref{fig04-prof-1} illustrate this difference.  
Here, the height range is chosen to be between 5 and 200 kpc from the galactic disk in order to avoid including the disk gas in our analysis.
The metal density in the {\tt Enzo} run decreases smoothly out to $z \sim 150$ kpc --- the edge of the metal-enriched halo gas --- at which point the density drops sharply.   
The {\tt Gad2} run without a gas halo in the IC shows only a few discrete points, indicating that only a small number of gas particles have been ejected into and remained in the halo. 
These discrete points have high outflow velocities, nearly $500 - 1000 \,\,{\rm km\,s^{-1}}$, as they do not have to move through any medium that decelerates the outflow.  
In contrast, once a gas halo is included in the IC ({\tt Gad2-GasHalo} run), the outflow velocity can reach only up to a few of $\sim 50 \,\,{\rm km\,s^{-1}}$.   
Finally, comparing the runs with a gas halo but with different thermal feedback energies --- {\tt Gad2-GasHalo/+TFB2/+TFB3} runs --- we find that the extent of metal-enriched gas is dictated by the feedback strength. 
The higher the feedback energy is, the faster the gas outflow becomes, enriching a larger volume of the halo.
The inclusion of the diffusion scheme does not affect the spatially averaged distribution of metals.

Figure \ref{fig04-prof-2} displays the enclosed metal masses as functions of the vertical height (left panel) and their time evolution (right panel).
We can again observe that the simulation with more feedback energy transports more metals from the disk to the halo.
In terms of the total metal mass in the halo, the mesh-based {\tt Enzo} run is most compatible with the {\tt Gad2-GasHalo+TFB2(+diff1)} run in both panels (as mentioned in Section \ref{sec:projection-fb} for Figure \ref{fig03-proj}).  
Without a gas halo ({\tt Gad2} run), the few metal-enriched gas particles rarely stay in the halo due to their high velocity, yielding an unrealistically metal-poor halo throughout the simulation.   

In the right panel of Figure \ref{fig04-prof-2}, the role of a gas halo in containing the SN ejecta is again illustrated.  
The metal mass in the region of $R_{200} < r < 2R_{200}$ in the {\tt Gad2} run (where $R_{200} = 205.5$ kpc) is shown with a thin dotted line, and this indicates that a few high-velocity SN ejecta particles have escaped the virial radius.  
They occasionally --- and unphysically --- reach thousands of kiloparsecs away from the galactic center. 
On the contrary, due to the presence of the halo gas, no ejected particle escapes the virial radius in all other runs. 
The gas halo, even when its density is negligible, imposes pressure on the gas outflows and decelerates them.
The confinement of metal-enriched outflows has been proposed by \cite{Ferrara2005}, who suggested that the gas surrounding the galactic disk exerts ram pressure on to the outflows so that the ejected metals are in a hot-diffuse phase.

Finally, in Figure \ref{fig04-prof-3}, we present the mass-weighted gas metallicity profiles in both the disk's radial direction and the vertical direction from the disk plane.
In the cylindrical radial direction (left panel), the metallicity in most runs is near the initial disk metallicity $Z_{\rm disk} = 0.02041$ (Section \ref{sec:IC}), and drops sharply at $r \sim25$ kpc --- the edge of the galactic disk. 
However, the (mass-weighted) metallicity is higher in all {\sc Gadget-2} runs in the galactic core than in the {\tt Enzo} run. 
It is because, in the particle-based simulations, metals tend to be locked in the dense region before they slowly disperse or diffuse into less dense regions. 
Meanwhile, the vertical metallicity profiles (right panel of Figure \ref{fig04-prof-3}) show a similar trend to the metal density profiles (the middle of Figure \ref{fig04-prof-1}) with one exception, the {\tt Gad2} run.
In the {\tt Gad2} run, the halo is insufficiently resolved with only a few high-velocity gas particles ejected from the metal-enriched star-forming regions; thus, the metal fraction in this region is not reliable.  

\vspace{1mm}

\subsection{Metal Distribution in The Density-Temperature Plane}\label{sec:pdf-fb}

We now investigate the metal distribution in the density-temperature phase space. 
In Figure \ref{fig05-pdf}, we draw the two-dimensional probability distribution functions (PDFs) of metal mass for various simulation setups with {\sc Enzo} and {\sc Gadget-2}. 
The one-dimensional projections along one of the axes --- i.e., density PDF and temperature PDF --- are shown in Figure \ref{fig06-1d-pdf} for more quantitative comparison.

\begin{figure*}[ht!]
\centering
\vspace{0mm}
\includegraphics[width=14.5cm]{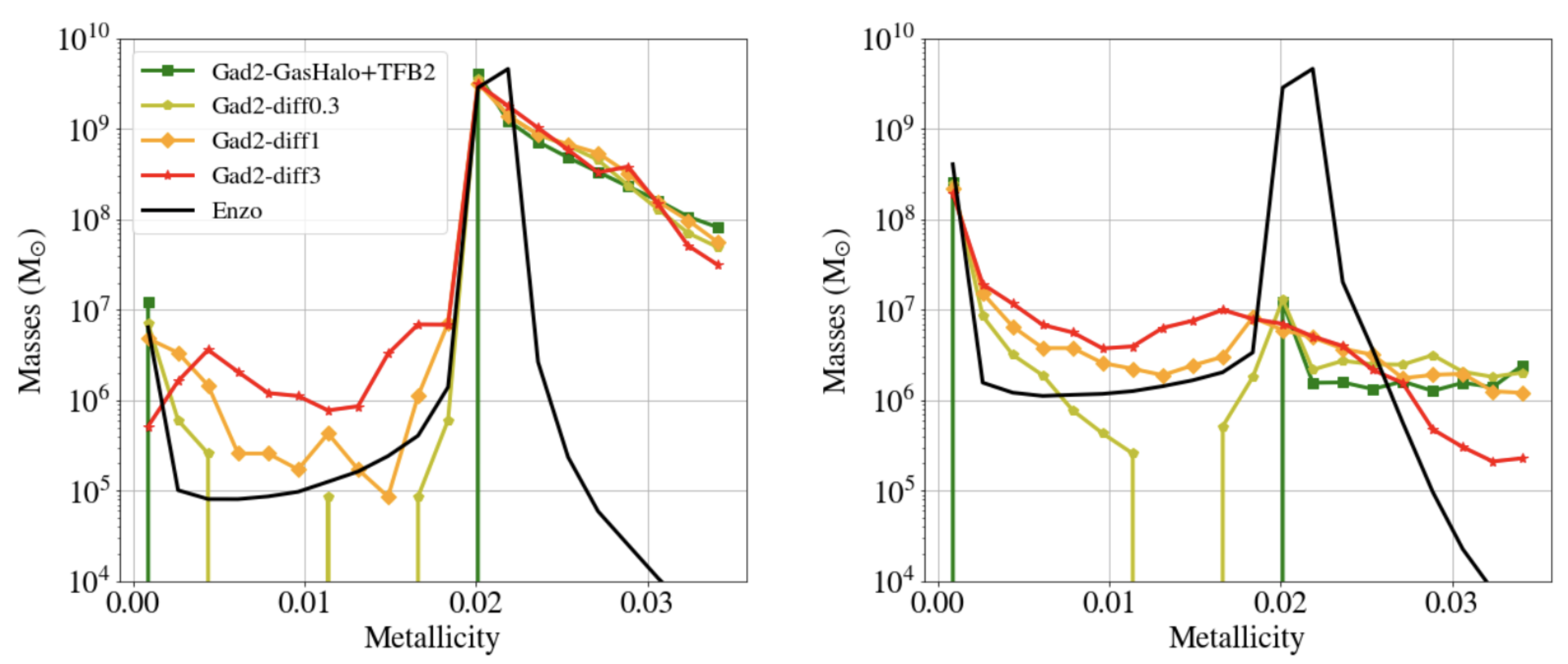}
\vspace{-1mm}
\caption{
The metallicity distribution functions (MDFs) for the disk gas ({\it left panel}) and the halo gas ({\it right panel}), in different simulation setups at 500 Myr. 
Here, the disk is defined as the region where the vertical distance from the disk plane $z$ is less than $5\,{\rm kpc}$. 
The halo is defined as the region where $ z > 5\,{\rm kpc}$ or the radial distance from the disk center $r > 60 \,{\rm kpc}$.
All {\sc Gadget-2} runs are with the gas halo and the thermal feedback energy of $2\times 10^{51}$ ergs per SN (i.e., {\tt GasHalo+TFB2}). 
Only with a sufficiently high metal diffusion coefficient $C_{\rm d}$ ($\geq 0.02$, i.e., {\tt Gad2-(GasHalo+TFB2)+diff1/+diff3} runs) can the particle-based simulations fill the domain of low-metallicity gas between $Z=0$ and $Z_{\rm disk}=0.02041$. 
See Table \ref{tab:2-runs} for the list of our simulations, and Section \ref{sec:MDF-diff} for more information on this figure.
}
\label{fig07-mdf}
\vspace{4mm}
\end{figure*}

In Figure \ref{fig05-pdf}, for all the runs considered, the majority of metal masses are on the thermal equilibrium curve stretching from $\sim10^4$ K to $\sim10^2$ K where cooling and heating rates are equal.
Three distinct phases of gas  ---  cold ($<10^3$ K), warm ($10^{3-5}$ K), and hot phase ($> 10^5$ K) --- are visible in all panels except {\tt Gad2}  (see also the right panel in Figure \ref{fig06-1d-pdf}).
The gas surrounding the disk (missing in the {\tt Gad2} run) is fed with hot gas particles expelled by SNe, and in turn, exerts pressure on the galactic outflow.
The dilute gas with varying entropy and pressure present in the {\tt Gad2} run, is now collapsed and confined to a constant pressure line between $10^{-30}$ and $10^{-27}\,{\rm g\,cm^{-3}}$ in the {\tt Enzo} and {\tt Gad2-GasHalo} runs.
This constant pressure phase develops via the pressure balance between the disk and the halo gas, and subsequently, a hot diluted halo is built.
The absence of a radiation channel via line emission at $\sim 10^6$ K thus creates a hot galactic halo in a thermodynamic equilibrium \citep{Ferrara2005}.
The metals in this hot phase gas can be hard to detect, potentially presenting a solution for the missing metal problem.

As we have discussed previously, the metal distribution in the galactic halo is greatly affected by the strength of thermal stellar feedback.
Comparing the runs with different thermal feedback energies --- {\tt Gad2-GasHalo/+TFB2/+TFB3} runs --- in Figure \ref{fig06-1d-pdf}, one can observe that the run with higher energy transports more metals to the hot-diffuse region.
In terms of the density and temperature PDFs in Figure  \ref{fig06-1d-pdf}, the {\tt Gad2-GasHalo+TFB2(+diff1)} run is the most compatible with the mesh-based {\tt Enzo} run (as discussed in Section \ref{sec:projection-fb} for Figure \ref{fig03-proj}, and in Section \ref{sec:profile-fb} for Figure \ref{fig04-prof-2}).\footnote{The extremely low density gas ($< 10^{-30}\,{\rm g\,cm}^{-3}$) displayed in the {\sc Gadget-2} runs is due to the near-empty region outside of the galactic virial radius.  
In contrast, in the {\sc Enzo} runs, the minimum gas density $n_{\rm H}=10^{-6}\,{\rm cm}^{-3}$ covers the entire simulation box outside the virial radius.}
The inclusion of the diffusion scheme does not significantly change the metal distribution in the density or temperature phase space.

\vspace{10mm}

\section{Metal Distribution In Halo: Dependence on Explicit Metal Diffusion Schemes}
\label{sec:5}

\vspace{1mm}

In this section, we investigate how the explicit turbulent metal diffusion scheme changes the metal content in a galaxy simulated with particle-based codes. 
We test different values for the metal diffusion coefficient ($C_{\rm d}$ in Section \ref{sec:diffusion}; see also Table \ref{tab:2-runs}) with a fixed stellar feedback model identical to {\tt Gad2-GasHalo+TFB2} that is shown to exhibit similar halo properties to the {\tt Enzo} run in Section \ref{sec:4}.

\begin{figure*}[ht!]
\vspace{0mm}
\centering
\includegraphics[width=18.6cm]{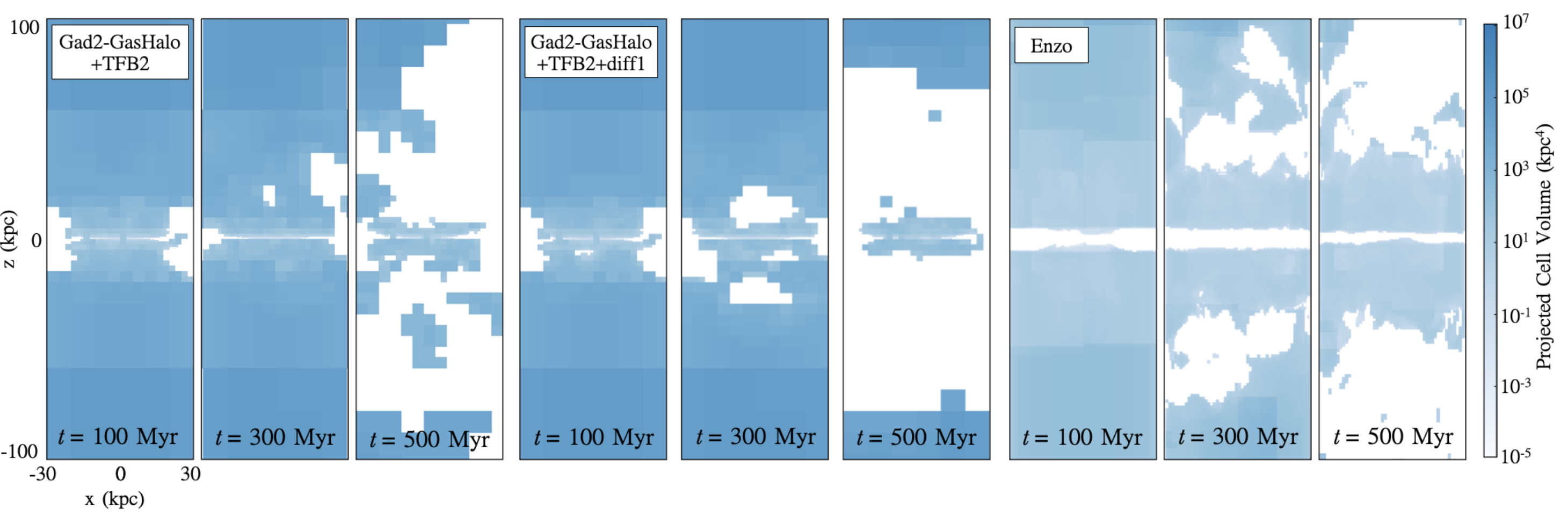}
\vspace{-5mm}
\caption{
Projected volume of metal-poor gas ($Z< 10^{-3}$) from the disk's edge-on angle  at 100, 300, and 500 Myr.
We perform the projection in the cylindrical volume of 30 kpc in radius and 200 kpc in height, centered on the galactic disk in the $z=0$ plane. 
The metal-enriched gas particles ejected by the stellar feedback gradually reduces the volume of metal-poor gas in time.  
By comparing the two particle-based simulations with and without the metal diffusion scheme, {\tt Gad2-GasHalo+TFB2} ({\it left panel}) and {\tt Gad2-GasHalo+TFB2+diff1} ({\it middle panel}), one can see that the scheme helps to enrich the halo with metals more effectively.   
See Section \ref{sec:5-2} for more information on this figure.
}
\label{fig08-pristine}
\vspace{4mm}
\end{figure*}

\begin{figure*}[ht]
\vspace{1mm}
\centering
\includegraphics[width=10.3cm]{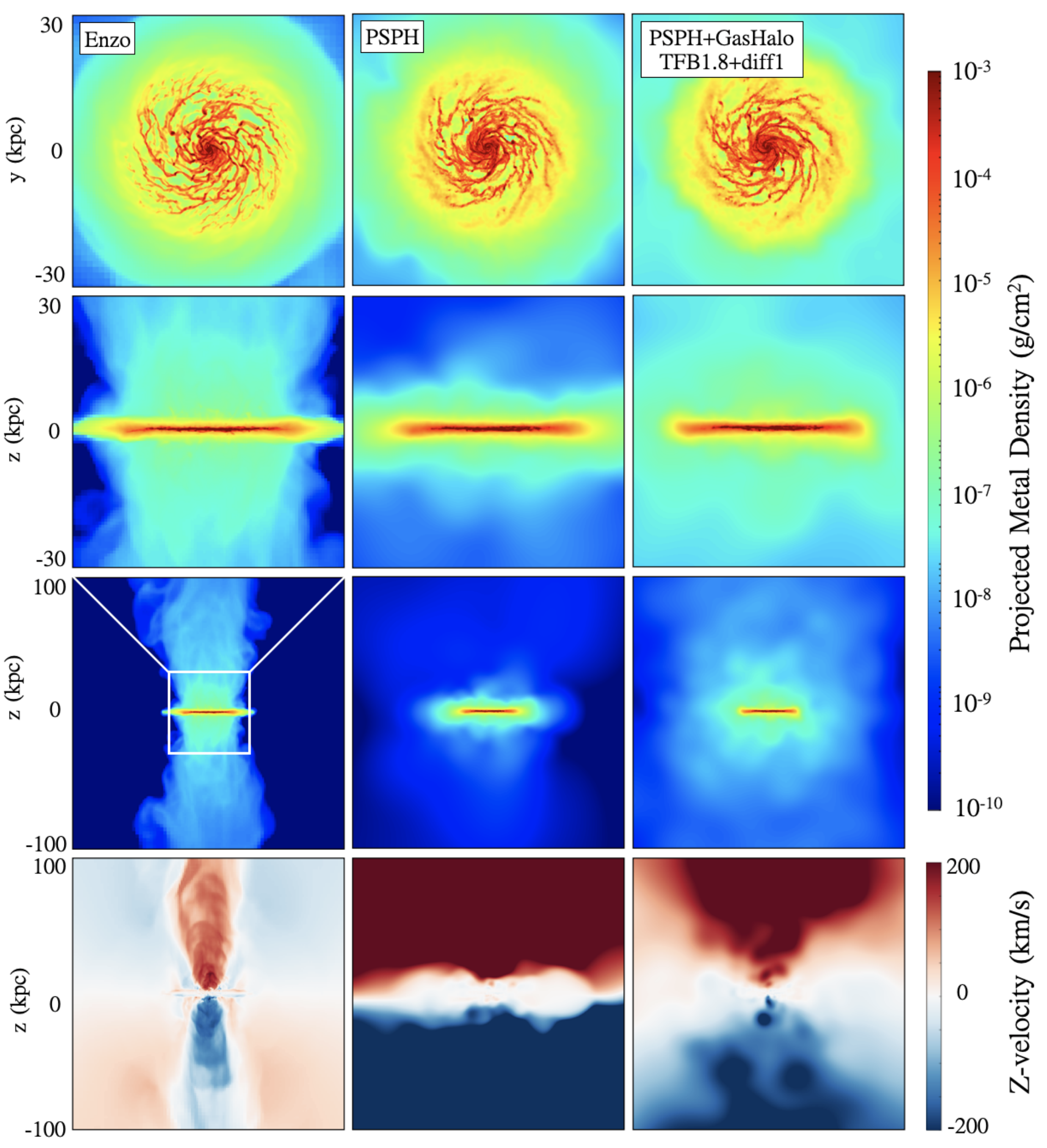}
\vspace{-1mm}
\caption{
$500$ Myr simulation snapshots similar to Figure \ref{fig03-proj}, but this time for the runs using the {\sc Gizmo-PSPH} code and our proposed prescription that makes a particle-based simulation compatible with the mesh-based {\tt Enzo} run --- stellar feedback with boosted thermal energy and the metal diffusion coefficient of $C_{\rm d}=0.02$.
The {\sc Gizmo-PSPH} simulation without the halo gas --- the {\tt PSPH} run (see Table \ref{tab:2-runs}); the same runtime condition as the particle-code runs in \citet{AGORA2016} --- substantially differs from the {\tt Enzo} run in the metal distribution inside the halo.  
In contrast, another {\sc Gizmo-PSPH} simulation, but this time with a gas halo, more feedback energy, and the explicit metal diffusion scheme --- the {\tt PSPH-GasHalo+TFB1.8+diff1} run  --- is compatible with the {\tt Enzo} run and the {\tt Gad2-GasHalo+TFB2+diff1} run in Figure \ref{fig03-proj}.  
See Table \ref{tab:2-runs} for the list of our simulations, and Section \ref{sec:6} for more information on this figure.
}
\label{fig09-best-proj}
\vspace{4mm}
\end{figure*}

\subsection{Metallicity Distribution Function}\label{sec:MDF-diff}

Figure \ref{fig07-mdf} presents the metallicity distribution functions (MDFs; metallicity PDFs) for the disk (left panel) and the halo gas (right panel).
We compare simulations with four diffusion coefficients, $C_{\rm d}=$ 0, 0.006, 0.02 and 0.06, labelled as {\tt Gad2-GasHalo+TFB2}, {\tt Gad2-(GasHalo+TFB2)+diff0.3}, {\tt Gad2-diff1}, and {\tt Gad2-diff3} run, respectively (see Section \ref{sec:diffusion} and Table \ref{tab:2-runs} for more information).
Since we have calibrated these runs so that the SFRs are similar (Figure \ref{fig02-sfr}), and their averaged metal profiles and metal masses are comparable (Figures \ref{fig04-prof-1}, \ref{fig04-prof-2} and \ref{fig06-1d-pdf}), we can conjecture that any discrepancy we see in the MDF is caused by varying the diffusion coefficient.  

In the {\tt Gad2-GasHalo+TFB2} run that does not include a diffusion scheme, the MDF (for both the disk and the halo) shows one sharp peak at $Z \simeq 0$ and another broader peak starting at $Z \simeq 0.02$.
The two values correspond to the initial metallicity values of the halo ($Z_{\rm halo}=10^{-6}\,Z_{\rm disk}$) and the disk ($Z_{\rm disk}=0.02041$), respectively. 
Unless an explicit diffusion scheme is used, gas particles residing only within the SN bubbles can acquire metals in a particle-based simulation.
The gas in the outer region, away from the star-forming regions, never receives or loses metals, making the MDF overly inhomogeneous.
Then, by increasing the diffusion strength, we can see that the gas metallicity is more evenly distributed.  
For example, in the {\tt Gad2-diff1/-diff3} runs, the gap between $Z=0$ and $Z_{\rm disk}$ is now filled, to a level of the {\tt Enzo} run.
The {\tt Gad2-diff1} run with $C_{\rm d}=0.02$ shows the most similar MDF to the {\tt Enzo} run,\footnote{As discussed in Section \ref{sec:diffusion}, the diffusion coefficient $C_{\rm d}=0.02$ is comparable to the value used by \citet{Shen2010}. 
It is, however, an order of magnitude higher than the one suggested by \cite{Escala2018}.} although none of the {\sc Gadget-2} runs produces the high peak at $Z=Z_{\rm disk}$  in {\tt Enzo}'s MDF for the halo.
In the meantime, the run with a weaker diffusion strength, the {\tt Gad2-diff0.3} run, fails to fully populate the domain between $Z=0$ and $Z_{\rm disk}$.

Therefore, one can conclude that an explicit metal diffusion scheme is essential in making the realistic low-metallicity gas in particle-based simulations. 
Our results suggest that the shape of an MDF is highly sensitive to the diffusion strength, both in the disk and in the halo. 
In particular, the diffusion scheme can help efficiently enrich the pristine gas at rest in the halo when it is penetrated by high-velocity metal-enriched outflows.
It is because, in the Smagorinsky-Lilly model, the flux of the metal field is proportional to its gradient and the velocity shear between the pockets of gas (see Section \ref{sec:diffusion}).
We however note that even the diffusion scheme has hard time mitigating the discrepancy at the high metallicity end ($Z > 0.02$ in Figure \ref{fig07-mdf}) between the {\sc Gadget-2} and {\sc Enzo} runs.  
This is related to another discrepancy seen in the left panel of Figure \ref{fig04-prof-3} at small cylindrical radii. 
The gas in the star-forming region remains metal-rich, and the metals therein are not readily dispersed even with the help of the explicit diffusion scheme (even with high $C_{\rm d}$).  

\begin{figure*}[ht]
\vspace{-2mm}
\centering
\includegraphics[width=17.8cm, height=5.9cm]{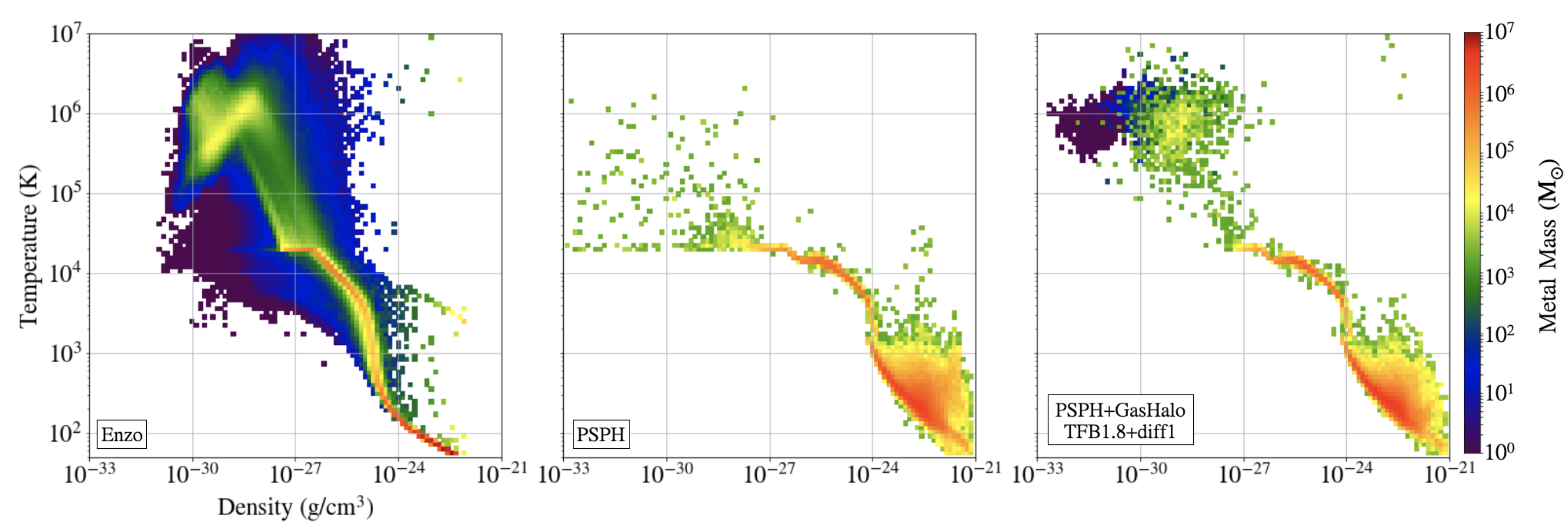}
\vspace{-2mm}
\caption{
Two-dimensional probability distribution functions (PDFs) of metal mass on the density-temperature plane at 500 Myr. 
The figure is similar to Figure \ref{fig05-pdf}, but this time for simulations using {\sc Gizmo-PSPH} and our proposed prescription that makes a particle-based simulation compatible with the mesh-based {\tt Enzo} run.
Only with a gas halo sufficiently resolved in the IC, more feedback energy, and the explicit metal diffusion (i.e., {\tt PSPH-GasHalo+TFB1.8+diff1}) can the particle-based simulations identify the hot diffuse gas around the disk at $[\sim 10^{-29} \,{\rm g\,cm^{-3}}, \,\sim10^6 \,{\rm K}]$.
See Section \ref{sec:6} for more information on this figure.
}
\label{fig10-best-pdf}
\vspace{4mm}
\end{figure*}

\subsection{Pockets of Metal-poor Gas} \label{sec:5-2}

We now look into the spatial distribution of pristine, metal-poor gas.  
Finding pockets of pristine gas --- if any --- that survived the metal contamination by its host galaxy has important implications in many studies in astrophysics: observations of the metal absorption lines in the CGM \citep[e.g.,][]{RocaFabrega19, Strawn2020}, the search for possible birthplaces for massive stars \citep[e.g.,][]{Turk2009, Sarmento2017}, and the search for massive black holes stemming (arguably) from the merging of massive stars \citep[e.g.,][]{Belczynski2010}, among others.  
Therefore, here we particularly focus on the metal-poor gas in the halo and investigate how its volume changes due to the inclusion of the explicit metal diffusion scheme.

Figure \ref{fig08-pristine} displays the projected volume of metal-poor gas ($Z< 10^{-3}$) from the disk's edge-on angle for different simulation setups at 100, 300, and 500 Myr. 
At 0 Myr, all metal-poor gas is by design only in the halo. 
As the galaxy evolves in time, the SN-driven winds make the metal-poor gas gradually disappear, starting from the regions closer to the disk.\footnote{Thin layers of metal-poor gas still appear above and below the galactic disk at 500 Myr in all of the runs shown in Figure \ref{fig08-pristine}.  
These layers of gas have been unaffected by the galactic wind because they are out of the wind's range of impact, off from the opening of the bipolar outflows.} 
Comparing the two {\sc Gadget-2} simulations with and without the metal diffusion scheme, {\tt Gad2-GasHalo+TFB2} and {\tt Gad2-GasHalo+TFB2+diff1} (left and middle panel), we find that the run with the scheme reduces the metal-poor gas more efficiently.   
As discussed in Section \ref{sec:MDF-diff}, the metal diffusion scheme in particle-based codes helps to redistribute the metals homogeneously in the halo and enriches a larger volume with metals.

\vspace{1mm}

Metal diffusion is a vital component in the process of galaxy formation.
In Section \ref{sec:5}, we have demonstrated that it has to be explicitly included in particle-based simulations to produce a realistic ISM and CGM in and around a simulated galaxy.  
Without considering the transport of metals via diffusion between gas particles, the MDF may become unreasonable (Figure \ref{fig07-mdf}), and pockets of unrealistically metal-poor gas may survive in the halo (Figure \ref{fig08-pristine}).

\vspace{1mm}

\section{Generalizing Our Findings In Another Particle-based Code}\label{sec:6}

\vspace{1mm}

In Sections \ref{sec:4} and \ref{sec:5}, using various metrics such as PDFs in density, temperature and metallicity, we have found that the {\tt Gad2-GasHalo+TFB2+diff1} run is the most compatible with the {\tt Enzo} run in reproducing its metal properties.  
The {\tt Gad2-GasHalo+TFB2+diff1} run features a sufficiently-resolved gas halo in the IC, stellar feedback with boosted thermal energy (twice the value used in the {\tt Enzo} run), and the metal diffusion with coefficient $C_{\rm d}= 0.02$.  
Before concluding our paper, in this section, we briefly test if our prescription for the {\sc Gadget-2} code is also applicable in the {\sc Gizmo-PSPH} code (see Section \ref{sec:particle-code}), and if our findings in {\sc Gadget-2} can be generalized in other particle-based simulations.
Readers should note that the authors never mean to imply that the {\tt  Enzo} run is a gold standard that all other simulations should match.  
We adopt the {\tt Enzo} run only as a reference while trying to find a setup that makes the mesh-based and particle-based codes behave in a similar fashion.

Figure \ref{fig09-best-proj} is similar to Figure \ref{fig03-proj}, but now the particle-based simulations are performed on the {\sc Gizmo-PSPH} code.  
The  {\tt PSPH} run in Figure \ref{fig09-best-proj} (2nd column; see also Table \ref{tab:2-runs}) behaves very similarly as the {\tt Gad2} run in Figure \ref{fig03-proj}, harboring an extremely metal-poor halo with only a few metal-enriched particles of very high outflow velocity.
Meanwhile, the {\tt PSPH-GasHalo+TFB1.8+diff1} run that utilizes our proposed prescription (3rd column; but with 10\% less energy than {\tt Gad2-GasHalo+TFB2+diff1}) shows a similar metal distribution and outflow velocity as the mesh-based {\tt Enzo} run.
The two-dimensional phase plot in Figure \ref{fig10-best-pdf} verifies the same trend.  
The {\tt PSPH} run in Figure \ref{fig10-best-pdf} (2nd panel) is similar to the {\tt Gad2} run in Figure \ref{fig05-pdf}, lacking the hot diffuse medium around the disk, unlike {\tt Enzo}.  
But with our proposed setup, the {\tt PSPH-GasHalo+TFB1.8+diff1} run (3rd panel) now features the hot metal-enriched medium just like the {\tt Gad2-GasHalo+TFB2+diff1} run in Figure \ref{fig05-pdf}.  
As a note, we have found that in the {\sc Gizmo-PSPH} simulation, slightly less (10\%) thermal feedback energy is required than in {\sc Gadget-2}, to best match the mesh-based {\tt Enzo} run.  
This small difference could be attributed to an inherent inter-code discrepancy between the pressure-energy formulation of SPH in {\sc Gizmo-PSPH} and the density-entropy formulation in {\sc Gadget-2}.

Based on these experiments, we argue that our proposed setup help to alleviate the discrepancy between mesh-based and particle-based codes previously reported in, e.g., \citet{AGORA2016}. 
Because our prescription is straightforward and relies only on the fundamental properties of SPH (see Section \ref{sec:particle-code} for more discussion), rather than a novel feature in any one code, we expect it to be widely applicable in many SPH codes.   
One may also argue that our criteria can be used to check if any particle-based simulation is robust and reproducible --- especially by a mesh-based code.  
For example, in a cosmological zoom-in simulation using a particle-based code, one may check if a galactic halo is resolved with a sufficient number of gas particles before analyzing its metal content or performing simulated metal line observations.

\vspace{1mm}

\section{Discussion and Conclusion}\label{sec:7}

\vspace{1mm}

Acquiring a realistic metal distribution in numerically-formed galaxies is vitally important, yet it is highly sensitive to the hydrodynamic schemes used and the diffusion model employed.
Indeed, the {\it AGORA} code comparison project has previously reported a nontrivial discrepancy in the metal distribution of an idealized galaxy simulation between mesh-based and particle-based codes \citep{AGORA2016}.
Following up on their observations, in this paper, we have investigated what causes the discrepancy and how it could be alleviated by changing the setup of a particle-based simulation.  
First, we have tested a modified IC for particle-based codes (Section \ref{sec:IC2}) that contains a large number of gas particles in the galactic halo to match the initial gas distribution of a mesh-based simulation.  
Then, we have examined the metal distributions in a suite of {\sc Gadget-2} simulations with different stellar feedback strengths and compare them with that of a {\sc Enzo} simulation (see Section \ref{sec:4}). 
We have also discussed the effect of an explicit metal diffusion scheme (Section \ref{sec:diffusion}) described in \citet{2018MNRAS.480..800H} and \citet{Escala2018},  and tested various coefficient values (Section \ref{sec:5}). 

We propose that, to alleviate the discrepancy in metal distributions between mesh-based and particle-based codes, the following three factors should be considered in a particle-based simulation:
$\,$ {\it (1) Sufficiently-resolved gas halo:}
Our study finds that a gas halo with density $n_{\rm H}=10^{-6}{\rm cm}^{-3}$ can provide enough pressure to contain the galactic outflows within the virial radius. 
A sufficient number of gas particles is needed in the halo to describe a well-resolved medium into which the energy and metals of the SN-driven outflows could be transferred.  
Consequently, the existence of the gas halo --- or the lack thereof --- heavily affects the metal distribution in it.
$\,$ {\it (2) Stellar feedback:} 
Stellar feedback is the main source of energy that maintains the hot-diffuse medium around the galactic disk. 
We find that the amount of metal-enriched gas and the metallicity profiles in the halo are dictated by the strength of thermal stellar feedback. 
Particle-based codes require approximately twice the thermal feedback energy as the mesh-based {\sc Enzo} code to produce compatible metal distributions in the halo.  
$\,$ {\it (3) Turbulent metal diffusion:}
We find that the explicit metal diffusion scheme based on turbulent mixing is essential to render a realistic low-metallicity gas in the galactic ISM and CGM. 
The shape of a metallicity PDF (or MDF) is highly sensitive to the strength of diffusion,  both in the disk and in the halo.   
The diffusion coefficient $C_{\rm d} = 0.02$ in a particle-based simulation provides the best match to a mesh-based {\sc Enzo} simulation.
Our proposed prescription combining the three factors above has been tested with two particle-based codes, {\sc Gadget-2} (Sections \ref{sec:4} to \ref{sec:5}) and {\sc Gizmo-PSPH} (Section \ref{sec:6}), and is generally applicable in many SPH codes.  

Even though the experiments reported in this paper have been performed with an idealized, isolated galaxy, our study offers a useful reference point for cosmological (zoom-in) simulations as well.  
For example, one may check if a galactic halo is sufficiently resolved in a particle-based simulation to make sure that any metal-related properties in the halo are reproducible by a mesh-based code (Section \ref{sec:6}).  
In the forthcoming paper, we will investigate the metal distribution inside the CGM in a full cosmological simulation with mesh-based and particle-based codes.  
We aim to examine how the predicted metal lines in the CGM and the pockets of pristine gas change as we adopt different hydrodynamic schemes (e.g., AMR vs. SPH vs. SPH+diffusion scheme). 
In addition, we will study the possibility of producing an extended metal-enriched CGM via a galaxy merger, inspired by the recent observations of widely extended or confined [CII] lines in high-$z$ galaxies \citep[e.g.,][]{Fujimoto2019, Ginolfi2020}.

\acknowledgments
The authors would like to thank Ena Choi, Myoungwon Jeon, Yongseok Jo, Woong-tae Kim, Kentaro Nagamine, Yves Revaz, Santi Roca-F\`abrega for insightful suggestions and helpful discussions. 
We also thank Volker Springel and Philip Hopkins for providing the public version of {\sc Gadget-2} and {\sc Gizmo}, respectively.
Ji-hoon Kim acknowledges support by Samsung Science and Technology Foundation under Project Number SSTF-BA1802-04.  
His work was also supported by the National Institute of Supercomputing and Network/Korea Institute of Science and Technology Information with supercomputing resources including technical support, grants KSC-2018-CRE-0052, KSC-2019-CRE-0163, and KSC-2020-CRE-0219.
The publicly available {\sc Enzo} and {\tt yt} codes used in this work are the products of collaborative efforts by many independent scientists from numerous institutions around the world.  
Their commitment to open science has helped make this work possible.  

%





\bibliography{citation}{}
\bibliographystyle{aasjournal}



\end{document}